\PassOptionsToPackage{unicode}{hyperref}
\PassOptionsToPackage{hyphens}{url}
\documentclass[
]{article}
\usepackage{xcolor}
\usepackage[margin=1in]{geometry}
\usepackage{amsmath,amssymb}
\usepackage{iftex}
\ifPDFTeX
  \usepackage[T1]{fontenc}
  \usepackage[utf8]{inputenc}
  \usepackage{textcomp} 
\else 
  \usepackage{unicode-math} 
  \defaultfontfeatures{Scale=MatchLowercase}
  \defaultfontfeatures[\rmfamily]{Ligatures=TeX,Scale=1}
\fi
\usepackage{lmodern}
\ifPDFTeX\else
\fi
\IfFileExists{upquote.sty}{\usepackage{upquote}}{}
\IfFileExists{microtype.sty}{
  \usepackage[]{microtype}
  \UseMicrotypeSet[protrusion]{basicmath} 
}{}
\makeatletter
\@ifundefined{KOMAClassName}{
  \IfFileExists{parskip.sty}{%
    \usepackage{parskip}
  }{
    \setlength{\parindent}{0pt}
    \setlength{\parskip}{6pt plus 2pt minus 1pt}}
}{
  \KOMAoptions{parskip=half}}
\makeatother
\usepackage{color}
\usepackage{fancyvrb}

\DefineVerbatimEnvironment{Highlighting}{Verbatim}{commandchars=\\\{\}}
\usepackage{framed}
\definecolor{shadecolor}{RGB}{248,248,248}
\newenvironment{Shaded}{\begin{snugshade}}{\end{snugshade}}

\newcommand{\AttributeTok}[1]{\textcolor[rgb]{0.13,0.29,0.53}{#1}}

\newcommand{\CommentTok}[1]{\textcolor[rgb]{0.56,0.35,0.01}{\textit{#1}}}

\newcommand{\DecValTok}[1]{\textcolor[rgb]{0.00,0.00,0.81}{#1}}
\newcommand{\DocumentationTok}[1]{\textcolor[rgb]{0.56,0.35,0.01}{\textbf{\textit{#1}}}}

\newcommand{\FunctionTok}[1]{\textcolor[rgb]{0.13,0.29,0.53}{\textbf{#1}}}

\newcommand{\NormalTok}[1]{#1}

\newcommand{\OtherTok}[1]{\textcolor[rgb]{0.56,0.35,0.01}{#1}}

\newcommand{\SpecialCharTok}[1]{\textcolor[rgb]{0.81,0.36,0.00}{\textbf{#1}}}

\newcommand{\StringTok}[1]{\textcolor[rgb]{0.31,0.60,0.02}{#1}}

\usepackage{graphicx}
\makeatletter
\newsavebox\pandoc@box
\newcommand*\pandocbounded[1]{
  \sbox\pandoc@box{#1}%
  \Gscale@div\@tempa{\textheight}{\dimexpr\ht\pandoc@box+\dp\pandoc@box\relax}%
  \Gscale@div\@tempb{\linewidth}{\wd\pandoc@box}%
  \ifdim\@tempb\p@<\@tempa\p@\let\@tempa\@tempb\fi
  \ifdim\@tempa\p@<\p@\scalebox{\@tempa}{\usebox\pandoc@box}%
  \else\usebox{\pandoc@box}%
  \fi%
}
\def\fps@figure{htbp}
\makeatother
\usepackage{listings}

\usepackage[numbers]{natbib}
\let\cite\citep

\usepackage{fancyhdr}

\usepackage{makecell}
\usepackage[percent]{overpic}

\usepackage{geometry}

\usepackage{setspace}
\usepackage{pdflscape} 
\usepackage{bookmark}
\IfFileExists{xurl.sty}{\usepackage{xurl}}{} 
\hypersetup{
  pdftitle={Non-parametric assessment of the calibration of individualized treatment effects},
  pdfauthor={Mohsen Sadatsafavi, Jeroen Hoogland, Thomas P.A. Debray, John Petkau},
  pdfkeywords={Clinical prediction models; Treatment benefit; Inference;
Hypothesis testing; Calibration},
  hidelinks,
  pdfcreator={LaTeX via pandoc}}

\title{Non-parametric assessment of the calibration of individualized
treatment effects}
\author{Mohsen Sadatsafavi, Jeroen Hoogland, Thomas P.A. Debray, John
Petkau}
\date{August 2026}

\begin{document}
\maketitle
\begin{abstract}
An important aspect of the performance of algorithms that predict
individualized treatment effects (ITE) is moderate calibration, i.e.,
the average treatment effect among individuals with predicted treatment
effect of z being equal to z. The assessment of moderate calibration is
challenging on two fronts: counterfactual responses are unobserved, and
quantifying the conditional response function for models that generate
continuous predicted values requires regularization. Perhaps because of
these challenges, there is currently no inferential method for the null
hypothesis that an ITE model is moderately calibrated in a population.
In this work, we propose non-parametric methods for the assessment of
moderate calibration of ITE models for binary outcomes using data from a
randomized trial. These methods simultaneously resolve both challenges,
resulting in novel graphical, numerical, and inferential methods for the
assessment of moderate calibration. The key idea is to formulate a
stochastic process for the cumulative prediction errors that obeys a
functional central limit theorem, enabling the use of the properties of
Brownian motion for asymptotic inference. We propose two approaches to
construct this process from a sample: a conditional approach that relies
on predicted risks (often an output of ITE models), and a marginal
approach based on replacing the cumulative conditional moments with
their marginal counterparts. Numerical simulations confirm the desirable
properties of both approaches and their ability to detect miscalibration
of different forms. We use a case study to provide suggestions on
graphical presentation and the interpretation of results. Moderate
calibration of predicted ITEs can be assessed without requiring
regularization techniques or making assumptions about the functional
form of treatment response. The accompanying cumulcalib R package
implements this method.
\end{abstract}

\let\thefootnote\relax\footnotetext{From Faculty of Pharmaceutical Sciences and Faculty of Medicine (MS) and Department of Statistics (JP), the University of British Columbia, Vancouver, Canada; Department of Epidemiology and Data Science, Amsterdam University Medical Centers, Amsterdam, The Netherlands (JH); Smart Data Analysis and Statistics B.V., Utrecht, the Netherlands, and University Medical Center Göttingen, Göttingen, Germany (TPAD)}

\let\thefootnote\relax\footnotetext{* Correspondence to Mohsen Sadatsafavi, Room 4110, Faculty of Pharmaceutical Sciences, 2405 Wesbrook Mall, Vancouver, British Columbia, V6T1Z3, Canada; email: mohsen.sadatsafavi@ubc.ca}

\clearpage

\section{Introduction}\label{introduction}

Due to their direct relevance to clinical decision-making, prediction
algorithms that produce quantitative estimates of individualized
treatment effect (ITE) have been the subject of much recent
development\cite{Kent2020PATH, Selby2025ITEREview, Desai2024ITEML}. A
common paradigm in medical practice is to offer treatment to patients at
high risk of untoward clinical outcomes, where the risk is quantified
via the application of prognostic markers or risk prediction
models\cite{Pauker1980}. Yet, evidence that these patients actually
benefit more than those with low outcome risk is often lacking. Indeed,
this risk-based approach towards treatment recommendation is often based
on strong and untested assumptions such as a monotonic association
between treatment effect and outcome risk. By directly producing an
estimate of treatment effect given an individual's salient
characteristics, ITE models enable more principled decision-making. An
example of such a model is the GRACE 3.0 ITE predictor for early
invasive management of non-ST-elevation acute coronary
syndrome\cite{Wenzl2025GRACE3}, developed using data from the VERDICT
trial. Another example is the ITE model for the effect of
oxygen-saturation targets on mortality in patient under intensive care,
based on data from the PILOT trial\cite{Buell2024}.

Before being implemented for patient care in a population, the
performance of an ITE model should be assessed using an independent
sample from that population --- a practice commonly referred to as
(external) validation\cite{Selby2024ITERev, Hoogland2024ITE}. A critical
aspect of ITE model performance in a validation study is calibration:
the ability of the model to generate predictions that are close to their
true counterparts\cite{Efthimiou2023ITEPErformance}. Poorly calibrated
ITE models may yield individualized predictions that are exaggerated, or
do not adequately reflect actual benefit. Consequently, they can result
in misguided clinical decisions by care providers and may mislead
patients in making treatment choices that are not aligned with their
preferences.

In the context of risk prediction models, calibration is defined at
different
levels\cite{PepeJanes2013, VanCalster2016calibrationHierarchy, Blanche2017WalleyPlot}.
The most basic form is calibration in the large (a.k.a. mean
calibration), which refers to the average predicted and average observed
risks being equal. Of more interest is assessing whether average
observed risk among all individuals with a given predicted risk is equal
to that predicted risk. This form of calibration is interchangeably
referred to as `weak' (e.g., by Pepe et al.\cite{PepeJanes2013} and
Blanche et al.\cite{Blanche2017WalleyPlot}) or `moderate' (e.g., by van
Calster et al.\cite{VanCalster2016calibrationHierarchy}) calibration.
Henceforth we use moderate calibration for consistency with our
directly-related previous work \cite{Sadatsafavi2024cumulcalib}. This
terminology is to contrast it with `strong' calibration, which refers to
the average observed and predicted risks being equal within every
subgroup defined by a combination of predictor
values\cite{PepeJanes2013, VanCalster2016calibrationHierarchy, Blanche2017WalleyPlot}.
However, it is argued that strong calibration is neither achievable nor
is it strictly desirable\cite{VanCalster2016calibrationHierarchy}. This
makes mean calibration (often the very first step in calibration
assessment) and moderate calibration of particular relevance. In
practice, mean calibration is assessed by comparing the average observed
and predicted risks, while moderate calibration is often assessed
visually using a flexible calibration plot, which depicts the
conditional mean of observed risk as a function of predicted risk. When
predicted risks are continuous with few or no ties, estimating this
conditional mean requires some form of smoothing or grouping. Scalar
metrics have also been proposed to summarize
it\cite{Harrell2015RegressionBook, VanHoorde2015ECI, Austin2019ICI}.
Because these visualizations and summary metrics rest on an estimated
conditional-mean function, their computation depends on tuning
parameters (e.g., the number of bins, or the LOESS bandwidth and
polynomial degree), and the choice of such parameters can have a
noticeable impact on the assessment.

Principles of calibration hierarchy can largely be extended to ITE
models\cite{Efthimiou2023ITEPErformance}. Mean calibration can be
examined by comparing the observed average treatment effect (ATE) and
the average of predicted effects. For moderate calibration, van Klaveren
et al.~quantified the difference between predicted and observed absolute
treatment effects in quartiles of predicted
effects\cite{vanKlaveren2019CalibForBenefit}. Maas et al.~proposed
applying local smoothing functions to match-based estimates of
conditional treatment effect, producing graphical presentations and
numerical metrics for the assessment of moderate
calibration\cite{Maas2023ITEPair}. Hoogland et al.~proposed fitting
logit models for outcome risk separately within control and treatment
groups, from which the calibration function for ITEs can be
derived\cite{Hoogland2024ITE}. However, these approaches require
grouping of data into bins, applying local smoothing methods, or using
parametric regression modeling.

Calibration can also be approached as a formal inference problem.
Inference on mean calibration for a risk prediction model entails
testing whether the average predicted risk equals the average observed
risk. Traditionally, inference on moderate calibration for risk
prediction models has been based on the Hosmer--Lemeshow test, albeit
this test is criticized for being sensitive to the inevitably arbitrary
number of groups (often set to 10)\cite{Hosmer1997GoFComparison}. For
ITE models, mean calibration can be tested using standard methods (e.g.,
testing for the equality of the average observed and predicted treatment
effect). However, to the best of our knowledge, there are currently no
inferential methods available for testing whether an ITE model is
moderately calibrated. This is an important methodological gap, as
achieving moderate calibration is an important criterion for many
clinical applications to proceed with the implementation of the model or
assessment of impact in further studies. Miscalibrated models might
result in loss of clinical benefit, as a treatment decision based on the
predicted ITE might be different from the decision had the true ITE been
known.

Recently, non-parametric approaches have been proposed for assessing
moderate calibration of risk prediction models based on the behavior of
cumulative prediction errors that do not require smoothing or grouping
of
observations\cite{Tygert2020Plots, ArrietaIbarra2022BM, Sadatsafavi2024cumulcalib}.
The core idea is that if the model is calibrated, the cumulative sum of
prediction errors should stay near zero, whereas for miscalibrated
models, the cumulative errors will tend to drift away from zero along
their path. These approaches yield a new class of visualizations,
summary metrics, and inferential tools for moderate calibration. In
particular, they enable formal inference on the null hypothesis that
``\emph{this model is moderately calibrated in this population}'', based
on asymptotic tests that rely on known properties of Brownian motion. It
would be appealing to extend this line of work to the assessment of
calibration of ITE models. This extension is challenging, as, unlike
risks, treatment benefits are not typically observable at the individual
level.

The purpose of this work is to propose non-parametric,
tuning-parameter-free approaches for the assessment of moderate
calibration of ITE models. Because of their causal underpinnings, ITE
models are often trained using data from randomized controlled trials
(RCTs). As such, we focus on validation studies based on RCT data, and,
as a first step, we consider binary outcomes. The rest of this
manuscript is structured as follows: after outlining the context, we
briefly review non-parametric calibration assessment for predicted
risks. We then extend this idea to ITE models by creating a stochastic
process that converges asymptotically to standard Brownian motion, which
is the basis for two approaches for constructing this process in the
sample. We conduct simulation studies examining the performance of tests
based on both approaches. These developments are exemplified in a case
study where we showcase the assessment of moderate calibration on the
cumulative domain. We conclude by discussing how these methods can be
added to the toolbox for evaluating ITEs, and suggest avenues for future
research, including their use in non-randomized settings.

\section{Methods}\label{methods}

\subsection{Notation and context}\label{notation-and-context}

We focus on binary outcomes (e.g., in-hospital mortality due to a severe
COVID-19 infection), and assume the outcome is an untoward event, and,
as such, treatment effect is in terms of absolute risk reduction. For
treatments that increase the probability of a desirable outcome, all
subsequent developments are applicable after relabeling the treatment
and control interventions. Let \(Y^{(0)}\) and \(Y^{(1)}\) be the
potential outcomes under control and treatment, respectively, both
taking a value in \{0,1\}. Let \(Y\) be the observed response, taking a
value in \{0,1\}. The treatment assignment indicator is \(a\) (0:
control, 1: treatment).

An ITE model is a deterministic function that for a given patient maps
their covariate pattern to a predicted treatment effect \(\delta\),
taking a value in (-1,1), which is claimed to be a good approximation
for \(\mathbb{E}(Y^{(0)}-Y^{(1)})\) among all individuals with the same
covariate pattern. The ITE model can be a regression-based or
machine-learning algorithm (as the functional form of the mapping
algorithm is not of direct relevance). Many ITE models are based on risk
prediction models that separately make predictions for outcome risk
under each treatment scenario, either from separate models (e.g., a T
learner) or a model with treatment as a predictor (a S
learner)\cite{Knzel2019MetaLearners}. These models therefore yield
counterfactual predictions: the expected outcome if treated and the
expected outcome if receiving the control treatment among all
individuals with the same covariate patterns. The predicted ITE is the
difference between these predicted values. For such models, by \(\pi\)
we refer to the `baseline' risks, which are claimed to be a good
approximation for \(\mathbb{E}(Y^{(0)})\). However, some modeling
approaches can generate predicted ITEs without modeling the underlying
risks\cite{Nie2020DirectITE, Tian2014}.

For our validation task, we assume we have access to an iid sample of
size \(n\) from a parallel-arm RCT. For the \(i\)th individual, we have
access to the following items: \(a_i\), representing treatment
assignment, the binary response variable \(Y_i\), and the predicted ITE
\(\delta_i\) which takes a value in (-1,1). For ITE models that also
generate predicted risks, we will also have access to the predicted
baseline risk \(\pi_i\) (which takes a value in (0,1)). We consider
\(\delta\)s and \(\pi\)s (if the ITE model also produces them) fixed at
their calculated values. While not strictly necessary, for ease of
derivations, we assume no ties in predicted benefits (i.e., at least one
of the predictors is a continuous variable).

Our goal is to construct a stochastic process \{S\}, that is, a process
with an implied time and an implied location component, based on the
cumulative sum of prediction errors that asymptotically converges to
Brownian motion, \(W(t)\), on the {[}0,1{]} time interval, if the ITE
model is moderately calibrated. This enables us to extend the previously
developed non-parametric approaches for calibration assessment of risk
prediction models to ITE models.

\subsection{Non-parametric approaches for moderate calibration of
predicted
risks}\label{non-parametric-approaches-for-moderate-calibration-of-predicted-risks}

In this section we briefly review non-parametric methods for the
assessment of moderate calibration of algorithms that predict outcome
risks given an individual's predictor pattern. These methods form the
basis of our subsequent developments for ITE models. To avoid
introducing new notation, in this section we assume none of the
individuals has received the treatment. As such, the baseline predicted
risks \(\boldsymbol{\pi}=(\pi_1,\pi_2,...,\pi_n)\) remain pertinent
predicted risks for the observed responses. For such predicted risks,
the null hypothesis that the model is moderately calibrated is defined
as \(\forall z: \mathbb{E}(Y | \pi=z)=z\). If a model generates
continuous predicted risks, estimating the conditional mean
\(\mathbb{E}(Y | \pi)\) requires regularization (binning or smoothing).
The core idea of non-parametric assessment is that one can instead
examine moderate calibration on the cumulative sum domain.

We start by indexing observations after ordering them from the small to
large values of the predicted risks \(\boldsymbol{\pi}\)s. We then
construct the scaled cumulative sums of the prediction errors. For the
\(k\)th observation, this is

\begin{equation}
  C_k = \frac{1}{n}\sum_{i=1}^k (Y_i-\pi_i), \text{for } k=1,2,...,n.
\end{equation}

\noindent It follows immediately from the definition of moderate
calibration that \(\forall k: \mathbb{E}(C_k)=0\) under the null
hypothesis. This expectation can be estimated from the data, without any
need for regularization.

If the model is not moderately calibrated, the \(C_k\)s tend to deviate
from zero along their path. Tygert proposed plotting \(C_k\) as a
function of \(k/n\) to inspect such deviation\cite{Tygert2020Plots}. The
degree of this deviation can be summarized into scalar metrics. One such
metric, proposed by Arrieta-Ibarra et al.\cite{ArrietaIbarra2022BM}, is
the scaled maximum absolute value of cumulative prediction errors:

\begin{equation}
  C^* = \max_{k} |C_k|.
\label{Cmax}
\end{equation}

\noindent This metric summarizes the degree of miscalibration, and can
be interpreted as the maximum absolute cumulative prediction error, as
an alternative to scalar summary metrics of miscalibration such as
\(E_{max}\) and the integrated calibration index
(ICI)\cite{Harrell2015RegressionBook, Austin2019ICI} - one whose
computation does not involve any tuning parameters (e.g., LOESS
bandwidth) and thus results in an objective, reproducible summary of
miscalibration.

Importantly, Arrieta-Ibarra et al.~argued that under the null hypothesis
that the risk prediction model is calibrated, the distribution of a
properly scaled version of \(C^*\) converges, as \(n \to \infty\), to
the distribution of the maximum absolute deviation of Brownian motion:
\(nC^*/\sqrt{\sum_{i=1}^n \pi_i(1-\pi_i)} \xrightarrow{d} \sup_{t \in [0,1]} |W(t)|\).
This is the test statistic of their proposed asymptotic test for
moderate calibration (henceforth referred to as the Brownian Motion
{[}BM{]} test).

Neither Tygert nor Arrieta-Ibarra et al.~explicitly discussed
constructing a stochastic process - a process with a time and a location
component - for prediction errors. The implied process in \(C_k\), with
time increments of \(1/n\) and location changes of \((Y-\pi)/n\), is
heteroscedastic, as the variances \(\pi(1-\pi)\) are variable but the
time increments are all equal. As such, while certain summaries of this
process converge to their counterparts for \(W(t)\) under \(H_0\), this
implied process does not achieve point-wise convergence to \(W(t)\).

Sadatsafavi and Petkau proposed turning these cumulative sums into a
stochastic process that converges at all points to \(W(t)\), thus
opening the door for taking full advantage of the properties of the
latter\cite{Sadatsafavi2024cumulcalib}. Let
\(s^2_k=\sum_{i=1}^k\pi_i(1-\pi_i)\) be the cumulative sum of variances
up to the \(k\)th observation. They defined the following standardized
process:

\begin{equation}
\{S\}=
\begin{cases}
  t_k = s^2_k/s^2_n & (\text{time values})\\
  
  S_k = \frac{nC_k}{s_n} & (\text{location values})
\end{cases},
\label{SProcess}
\end{equation}

\noindent (with \(t_0=0\) and \(S_0=0\)). The time increments are scaled
to sum to 1, and the variances of the location changes are scaled to be
equal to the time increments. This process is, under \(H_0\), a
martingale with independent and strictly bounded increments, with
\(s^2_n \to \infty\) as \(n \to \infty\). Thus, Brown's martingale CLT
establishes that the continuous process based on linear interpolation of
successive (\(t_k, S_k\)) points converges weakly to \(W(t)\) on the
{[}0,1{]} time interval\cite{Brown1971MartingaleCLT}.

As an immediate application of this convergence, they proposed a
two-part `bridge' test. This test is motivated by the following
observations. First, under \(H_0\),
\(S_n \xrightarrow{d} W(1)\sim\mbox{Normal}(0,1)\). Thus, \(S_n\), the
test statistic for the first part, is a z-score for mean calibration, a
necessary condition for moderate calibration. Second, the path of a
`bridged' Brownian motion, which is defined as the difference between
\(W(t)\) and the line that connects its endpoints (also known as the
Brownian bridge\cite{Chow2009BrownianBridge}) is independent of its
terminal value: \(\forall t \in [0,1): W(t)-tW(1) \perp W(1)\). As
miscalibrated models tend to generate large deviations from this path,
examining the maximum absolute deviation of the bridged random walk
provides an opportunity, independently of \(S_n\), for assessing
moderate calibration. The test statistic for the second part is thus
\(S^*=\max_k(|S_k-t_kS_n|)\), with
\(S^* \xrightarrow{d}\sup_{t\in[0,1]}|W(t)-tW(1)|\) under \(H_0\). The
distribution of the maximum absolute deviation of the Brownian bridge is
the Kolmogorov distribution, the asymptotic null distribution of the
two-sided one-sample Kolmogorov-Smirnov test
statistic\cite{Smirnov1948, Doob1949}. The p-values of each part are
combined via Fisher's method to generate a unified p-value for moderate
calibration. Simulations confirmed the intuition that, being based on
two independent statistics, the bridge test generally has higher power
than the BM test. The authors also proposed graphical visualization of
the homoscedastic \{S\} process as an alternative to the presentation of
the heteroscedastic scaled cumulative calibration errors (C).

\begin{figure}[h!]
    \centering
    \caption{Example of the \{S\} process for a calibrated model$^*$ (A), a model that underestimates the actual outcome (B), and an overfitted model that makes extreme predictions (C)}
    
    \begin{tabular}{ccc}
        \includegraphics[width=0.31\textwidth]{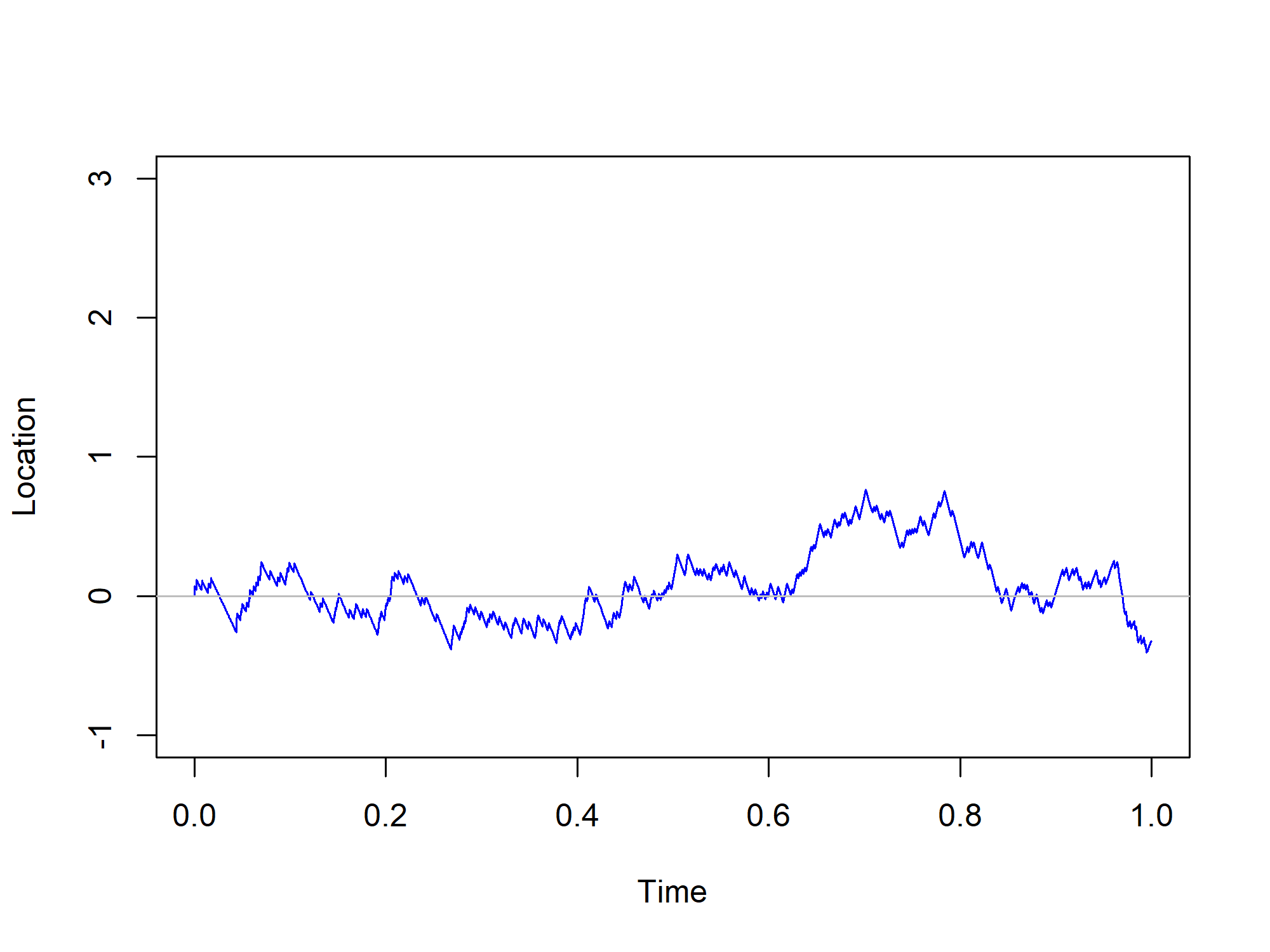} &
        \includegraphics[width=0.31\textwidth]{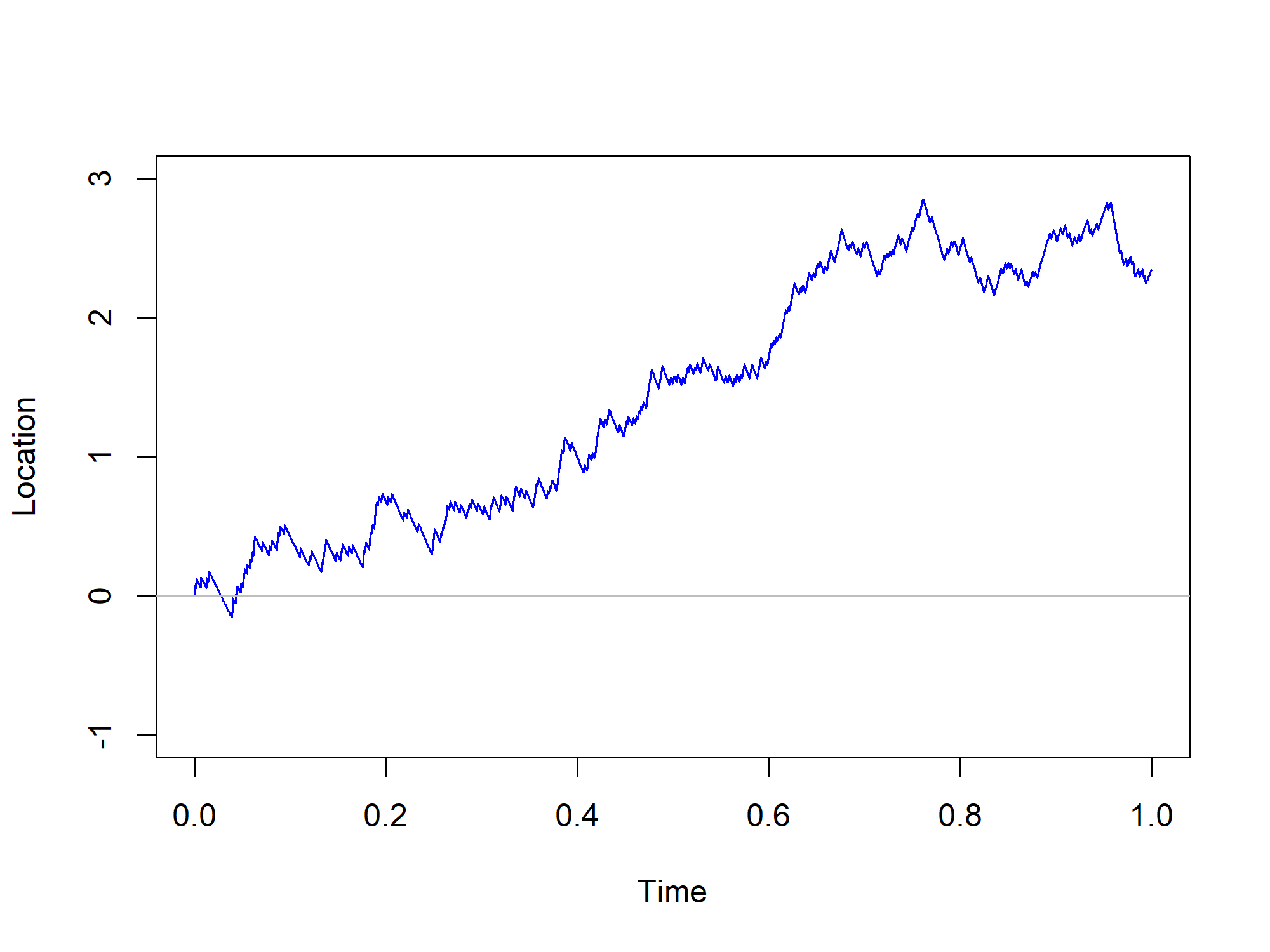} &
        \includegraphics[width=0.31\textwidth]{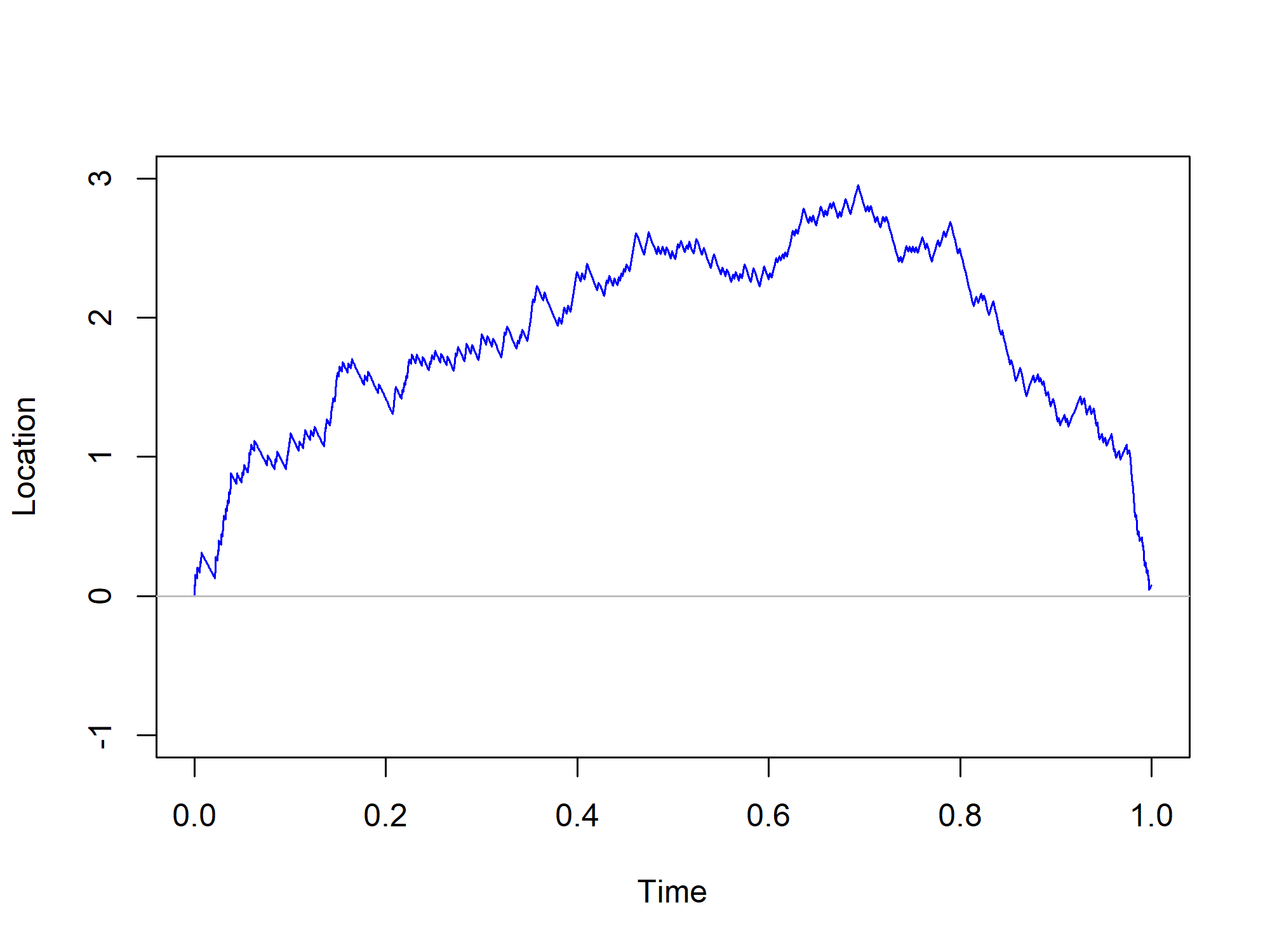} \\
        \small\bfseries (A) & \small\bfseries (B) & \small\bfseries (C)
    \end{tabular}

    \vspace{1em}
    
    \begin{minipage}{\textwidth}
        \footnotesize
        \raggedright 
        * The construction and interpretation of the {S} process are the same for risk-prediction and ITE models: the qualitative reading of the trajectory as random fluctuation (A), systematic drift (B), or an inverse-U departure (C), carries over directly. \par
        \medskip
        The true model is $\mbox{logit}(P(Y=1))=-1+X$ with $X\sim \mbox{Normal}(0,1)$. The three prediction models are as follows: A: same as the true model; B: $\mbox{logit}(P(Y=1))=-1.2+X$; C: $\mbox{logit}(P(Y=1))=-1.2+1.2X$. Results are from a random dataset of 1,000 observations. Time and Location axes refer to their corresponding elements in Equation \ref{SProcess}. \par
        \medskip
        Note the random variation of cumulative errors in Panel A (typical of a calibrated model), the systematic increase in panel B (indicating a systematic underestimation of risks), and the inverse U-shape pattern in panel C (indicating an over-dispersed model that makes overly extreme predictions --- underestimating risks when they are low and overestimating them when they are high). \par
    \end{minipage}
    \label{fig:stylized_rw}
\end{figure}

Figure \ref{fig:stylized_rw} provides examples of the visualization of
the \{S\} process and typical miscalibration patterns. While this
exemplary process is built around a risk prediction model, the
interpretation of the process will be identical for ITE models.

\subsection{Non-parametric approaches towards calibration of ITE
models}\label{non-parametric-approaches-towards-calibration-of-ite-models}

Our main purpose is to generalize the above-mentioned developments to
the assessment of the calibration of ITE models. Before proceeding to
main developments, we note that for ITE models that predict risks under
control and treatment, predicted ITEs are the difference between these
predicted risks. If predicted risks under both treatment conditions are
to be communicated to the patient and be used in decision-making, it is
natural to demand that they both be moderately
calibrated\cite{Hoogland2024ITE}. For this purpose, one can define a
compound null hypothesis as
\(H_{0A}: \forall z: \mathbb{E}(Y^{(0)}|\pi=z)=z\) and
\(H_{0B}: \forall z: \mathbb{E}(Y^{(1)}|\pi=z)=z-\delta\). In the
context of a randomized trial, and under classical assumptions for the
potential outcomes framework (consistency, exchangeability, positivity,
and no interference\cite{Hernan2025CausalInference}), these hypotheses
can also be written as
\(H_{0A} : \forall z : \mathbb{E}(Y |a = 0, \pi = z) = z\) and
\(H_{0B} : \forall z : \mathbb{E}(Y |a = 1, \pi = z) = z - \delta\).
Because the subgroups of treated and control individuals are independent
of each other, this compound null hypothesis can be tested via separate
non-parametric assessment of the calibration of predicted risks within
each treatment group. A unified p-value can be generated via Fisher's
method. A visualization of this approach would involve plotting the
\{S\} processes within each treatment group. However, this approach
fails to focus the attention on the component of interest here which is
ITE calibration. Indeed, the ITEs can remain calibrated despite
miscalibrated predicted risks. The visualization of cumulative predicted
risk errors also does not provide an opportunity for a direct assessment
of the cumulative prediction errors for ITEs. Further, this approach is
not applicable to ITE models that do not output predicted risks. As
such, for the rest of this work we study stochastic processes that focus
on the calibration of predicted ITEs.

In this context, the null hypothesis that an ITE model is moderately
calibrated can be formulated as

\begin{gather}
H_0: \forall z: \mathbb{E}(Y|a=0,\delta=z)-\mathbb{E}(Y|a=1,\delta=z)=z.
\label{H0}
\end{gather}

\subsubsection{A stochastic process for cumulative
ITEs}\label{a-stochastic-process-for-cumulative-ites}

Our goal is to create a standardized \{S\} process for cumulative ITEs
with weak convergence to \(W(t)\), enabling the use of the
above-mentioned visualizations, summary indices, and inference for the
assessment of ITE models.

To transform the assessment domain to cumulative prediction errors, we
start by indexing the observations after ordering the sample from small
to large values of predicted ITEs, such that
\(\forall i<j: \delta_i < \delta_j\). Had we access to both potential
responses at the individual level, we would be able to study, similarly
to the approach used for predicted risks, an appropriately standardized
version of cumulative prediction errors
\(\sum_{i=1}^k ((Y^{(0)}_i-Y^{(1)}_i)-\delta_i)\). However, one of
\(Y^{(0)}\) and \(Y^{(1)}\) is counterfactual and is therefore
unobserved.

Our core idea is to work with the following estimator in lieu of
observed cumulative treatment effects:

\begin{gather}
B_k = k \times[\text{average risk reduction due to treatment in the first $k$ subjects}]= \\
k\left( \frac{\sum_{i=1}^k(1-a_i)Y_i}{\sum_{i=1}^k(1-a_i)}-\frac{\sum_{i=1}^k a_iY_i}{\sum_{i=1}^k a_i} \right),
\nonumber
\label{BProcess}
\end{gather}

\noindent for \(k=1,2,3,...,n\). By convention we define \(0/0=0\). This
process constructs an estimator of cumulative ITEs as \(k\) times the
estimate of the ATE among the first \(k\) subjects.

Based on this estimator, we construct a martingale to which a functional
CLT can be applied. To this end, for the \(k\)th observation, the
conditional expected values and conditional variances of the increments
given the history of the cumulative sum up to the \(k\)th observation
are required\cite{HallHeyde2014martingaleCLT}. Let \(\mathcal{F}_k\)
denote all the information embedded in the history up to and including
the \(k\)th observation. The increments (differences) \(D_k\) and their
conditional expectations (\(\mu_k\)) are defined as (we set \(B_0=0\)):

\begin{equation}
  D_k=B_k-B_{k-1}, \qquad \mu_k=\mathbb{E}(B_k-B_{k-1}|\mathcal{F}_{k-1}).
\end{equation}

\noindent The equations for \(D_k\) and \(\mu_k\) are provided in
Appendix 1. Of particular interest are the centered increments:

\begin{equation}
  D_k-\mu_k=k\left[(1-a_k)\frac{Y_k-\pi_k^*}{\sum_1^{k}(1-a_i)}-a_k\frac{Y_k-\pi_k^*+\delta_k^*}{\sum_{i=1}^k a_i}\right],
\label{condinc}
\end{equation}

\noindent where \(\pi^*_k=\mathbb{E}(Y|\delta=\delta_k, a=0)\) and
\(\delta^*_k=\mathbb{E}(Y|\delta=\delta_k, a=0)-\mathbb{E}(Y|\delta=\delta_k, a=1)\)
are, respectively, the expected baseline risk and expected ITE among
individuals with the predicted ITE equal to \(\delta_k\).

The scaled cumulative sum of the centered increments thus becomes:

\begin{equation}
  C_k = \frac{1}{n}\sum_{i=1}^k (D_i-\mu_i).
\end{equation}

\noindent To create the \{S\} process for the application of Brown's
martingale CLT\cite{Brown1971MartingaleCLT}, we need to standardize the
time differences according to the cumulative conditional (on the
history) variances. The individual conditional variances are

\begin{equation}
  \sigma^2_k=\mbox{Var}(D_k|\mathcal{F}_{k-1})=
  k^2\left[(1-a_k)\frac{\pi_k^*(1-\pi_k^*)}{(\sum_{i=1}^{k}(1-a_i))^2}+
  a_k\frac{(\pi_k^*-\delta_k^*)(1-\pi_k^*+\delta_k^*)}{(\sum_{i=1}^{k}a_i)^2}\right],
\label{condvar}
\end{equation}

\noindent and the total conditional variance at point \(k\) is

\begin{equation}
  s^2_k=\sum_{i=1}^k\sigma^2_i.
\end{equation}

\noindent Putting all these together, we construct the stochastic
process of interest (\{S\}) as in equation \ref{SProcess}, with a time
component of \(t_k = s^2_k/s^2_n\) and a location component of
\(S_k = \frac{nC_k}{s_n}\).

Similar to the \{S\} process for risks, time steps in this process sum
to one, and the conditional variances of the increments are scaled to be
equal to the time increments. We note that the increments in this
process (\(D_k-\mu_k\)) are independent, but unlike their counterparts
for risks, are not strictly bounded - as \(k/\sum_{i=1}^ka_i\) and
\(k/\sum_{i=1}^k(1-a_i)\) can take arbitrarily large values. The
application of the functional CLT is no longer straightforward. However,
in Appendix 2, we show that under general regularity conditions for
parallel-armed RCTs, this process does satisfy the Lindeberg condition
for martingales, and Brown's martingale CLT establishes that the
continuous process based on linearly interpolating successive points of
\{S\} converges weakly to \(W(t)\) on the {[}0,1{]} time interval.

The conditional moments of the above process depend on calibrated ITEs
and baseline risks. Indeed, under \(H_0\), our claim is that
\(\delta=\delta^*\), and so we replace \(\delta^*\) with \(\delta\) in
equations \ref{condinc} and \ref{condvar}. But a practical challenge
remains which is the need for calibrated baseline risks \(\pi^*\). Two
approaches are developed to address this issue. The first uses predicted
risks in place of \(\pi^*\). The second replaces cumulative conditional
moments of the above process with their marginal counterparts, which do
not require knowing the true risks.

\subsubsection{Approach 1: A conditional (on predicted risks)
process}\label{approach-1-a-conditional-on-predicted-risks-process}

For ITE models that produce predicted risks, an obvious solution is to
replace \(\pi^*\)s with their predicted counterparts. This results in an
approach that is conditional on the assumption that the predicted
baseline risks are well-calibrated within strata of predicted ITEs. One
can then generate, just like for risk, the graph of \(C_k\) as a
function of \(k/n\). As well, \(C_n\) is a measure of mean calibration,
albeit not the conventional one based on comparing the observed and
predicted ATE (\(B_n/n-\sum_{i=1}^n\delta_i/n\)). Further, similarly as
for risks, \(C^*=\max_k |C_k|\) is a scalar metric that captures the
degree of miscalibration: miscalibrated ITE models tend to generate
larger \(C^*\) values compared with calibrated ones.

The main application of this approach arises in models that construct
predicted ITEs from predicted risks for treatment and control, where the
calibration of predicted risks is first evaluated, and the risk
prediction component is revised if necessary to ensure calibration. A
related example is when a previously validated risk prediction model is
combined with an estimated treatment effect from a clinical trial to
generate ITE predictions. In such cases, it is reasonable to condition
inference about the ITEs on the premise that the predicted risks are
calibrated. However, this approach does not apply to ITE models that do
not generate predicted risks.

\subsubsection{Approach 2: An approximate marginal (independent of
predicted risks)
process}\label{approach-2-an-approximate-marginal-independent-of-predicted-risks-process}

If baseline risks are not available or cannot be trusted as being
calibrated, it will not be possible to construct a martingale whose
asymptotic behavior can be fully specified. This is because the
conditional expected values (\(\mu_k\)) and variances (\(s^2_k\)) of the
increments depend on baseline risks. However, one can approximate the
cumulative sum of conditional expected values and variances using their
marginal counterparts, whose values can be obtained without knowledge of
the baseline risks.

Under \(H_0\), as we claim \(\delta=\delta^*\), at point \(k\) the
expected number of avoided outcomes is \(\sum_{i=1}^k \delta_i\);
therefore, the expected value of \(B_k-B_{k-1}\) can be approximated by
\(\tilde{\mu}_k = \delta_k\), and thus we approximate the scaled
cumulative prediction errors as:

\begin{equation}
  \tilde {C}_k = \frac{1}{n}(B_k-\sum_{i=1}^k \delta_i).
\end{equation}

\noindent As well, we replace the cumulative sum of conditional
variances up to point \(k\) with the total marginal variance of \(B\) at
this point:

\begin{equation}
  \tilde{s}_k^2 = \mbox{Var}(B_k) = k^2\left( \frac{p_0(1-p_0)}{\sum_{i=1}^k (1-a_i)} + \frac{p_1(1-p_1)}{\sum_{i=1}^k a_i} \right),
\end{equation}

\noindent where
\(p_0=\frac{\sum_{i=1}^k (1-a_i)Y_i}{\sum_{i=1}^k (1-a_i)}\),
\(p_1=\frac{\sum_{i=1}^k a_i Y_i}{\sum_{i=1}^k a_i}\) are the sample
estimates, up to the \(k\)th observation, of the outcome risk in each
arm. The \{S\} process is defined based on these \(\tilde C\) and
\(\tilde{s}^2\) values replacing their counterparts in equation
\ref{SProcess}.

\begin{figure}[h!]
    \centering
    \caption{Cumulative conditional versus marginal expected values (top-left), cumulative conditional versus marginal variances (top-right), and corresponding \{S\} processes. Black: Conditional process; Blue: Marginal process. These are from an ITE model of the form $\mbox{logit}(\pi_a(x))=0+0.25x+-0.5a + 0.25xa$ with $x$ being the single predictor with a standard Normal distribution, and a simulated sample of size 2,000.}
    \includegraphics[scale=0.6]{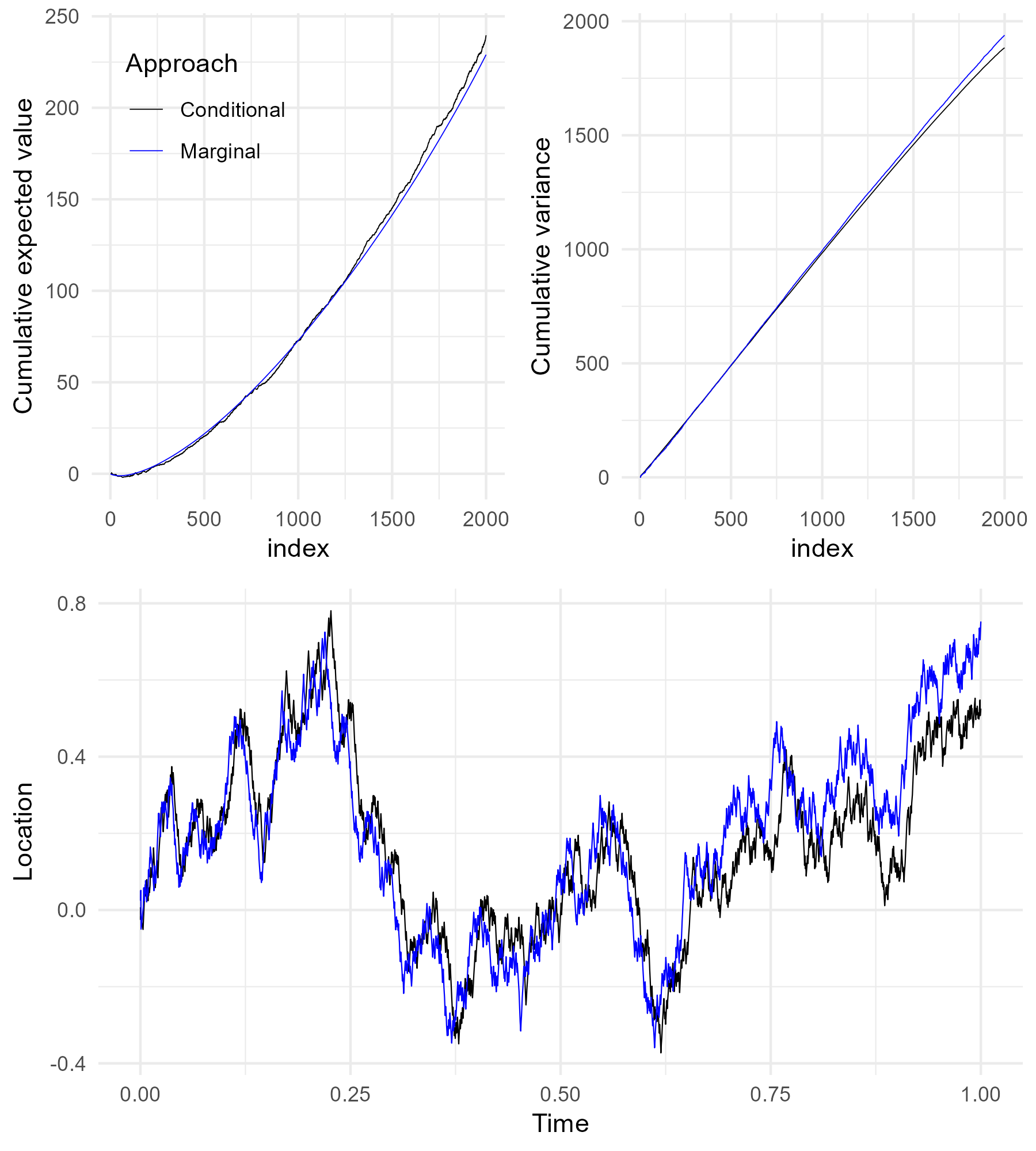}
    \label{fig:cond_v_marg}
\end{figure}

We note that the asymptotic behavior of this marginal process is not
fully supported by theory, and this approach remains heuristic in
character. But our intuition is supported by the proximity of the
resulting stochastic processes, and is confirmed in a more substantial
fashion by the simulation studies of the next section. A visual example
is provided in Figure \ref{fig:cond_v_marg}. The top panels juxtapose
the cumulative conditional and marginal moments from an exemplary
sample. The bottom panel demonstrates the corresponding \{S\} processes.
The general features of the processes are similar but shifted slightly
both horizontally, due to discrepancies between variance terms, and
vertically, due to discrepancies between expected values. We emphasize
that this marginal process only approximates the stochastic process of
interest. Specifically, while the marginal variances generally tend to
increase, we might occasionally see \(\tilde{s}^2_k<\tilde{s}^2_{k-1}\),
resulting in occasional negative time increments. On the other hand, one
attractive feature of this process is that its terminal value
(\(\tilde{C}_n\)) is the conventional measure of mean calibration: the
difference between the conventional estimate of ATE and the sample
average of predicted ITEs.

\section{Simulation studies}\label{simulation-studies}

In this section we report on proof-of-concept simulations that examine
the behavior of the proposed approaches. Three sets of simulations were
performed. In the first set we assessed the null distribution of
p-values, and in the second and third sets we investigated the power of
these tests in detecting various forms of miscalibration. Each
simulation scenario was run 10,000 times, a value that results in Monte
Carlo standard errors of at most 0.005 for the probability of rejecting
\(H_0\) at 5\% significance level (regardless of the actual rejection
rate); and SE\textless0.003 for the average p-value, thus making the
first two digits of rejection probabilities and average p-values
significant. We tested both the one-part BM test and the two-part bridge
test for both conditional and marginal approaches (four tests in total).
We did not compare these tests against any other test, as we are not
aware of any other inferential method for our null hypothesis.

All simulation scenarios involve a reference ITE model of the form
(\(\pi_a\) indicates predicted risks under treatment \(a\)):

\[\mbox{logit}(\pi_a(x))=\beta_0+\beta_x x+\beta_aa+\beta_{xa}xa,\]

\noindent with \(x\) the single predictor, \(a\) the binary treatment
variable, and \(Y\) the binary outcome. Predicted ITEs from this model
are generated as \(\delta(x)=\pi_0(x)-\pi_1(x)\).

For the first set of simulations, we generated \(n\) simulated
individuals given independent draws \(x_i \sim \mbox{Normal}(0,1)\),
\(a_i \sim \mbox{Bernoulli(0.5)}\), \(\pi_i\) from the reference
prediction model, and \(Y_i \sim \mbox{Bernoulli}(\pi_i)\). In a
factorial design, we changed the parameters of the reference model as
follows: \(n \in \{500,5000\}\), \(\beta_0 \in \{-1,1\}\),
\(\beta_x \in \{0,0.25\}\), \(\beta_a \in \{-0.5,-0.25\}\), and
\(\beta_{xa} \in \{0,0.25\}\).

Table \ref{tab:h0_reject} reports, for each of the four tests across all
32 null scenarios, the mean p-value and the empirical rejection rate at
the 0.05 significance level. As the distribution of p-values should be
uniform under \(H_0\), the mean p-value should be close to 0.5 and the
rejection rate close to its nominal level. This is the case for all four
tests, even in scenarios with relatively small samples for ITE model
validation.

\begin{landscape}
\begin{table}[p]
\centering
\caption{Mean p-value and empirical rejection rate at the 0.05 significance level under $H_0$ across the null scenarios, for the four tests. Under $H_0$ the p-values are uniformly distributed, so the mean p-value is expected to be near 0.5 and the rejection rate near 0.05. Each cell is based on 10,000 simulation replications.$^*$}
\resizebox{\linewidth}{!}{%
\begin{tabular}{rrrrrcccccccc}
\hline
\multicolumn{5}{c}{Scenario} & \multicolumn{4}{c}{Mean p-value} & \multicolumn{4}{c}{Rejection rate at $\alpha=0.05$} \\ 
\cline{6-9}\cline{10-13}
$n$ & $\beta_0$ & $\beta_x$ & $\beta_a$ & $\beta_{xa}$ & BM-conditional & Bridge-conditional & BM-marginal & Bridge-marginal & BM-conditional & Bridge-conditional & BM-marginal & Bridge-marginal \\ 
\hline
500 & -1.00 & 0.00 & -0.50 & 0.00 & 0.512 & 0.516 & 0.509 & 0.511 & 0.047 & 0.048 & 0.049 & 0.050 \\
500 & -1.00 & 0.00 & -0.50 & 0.25 & 0.520 & 0.522 & 0.517 & 0.517 & 0.045 & 0.043 & 0.047 & 0.046 \\
500 & -1.00 & 0.00 & -0.25 & 0.00 & 0.512 & 0.515 & 0.510 & 0.511 & 0.047 & 0.045 & 0.048 & 0.046 \\
500 & -1.00 & 0.00 & -0.25 & 0.25 & 0.509 & 0.512 & 0.506 & 0.507 & 0.047 & 0.045 & 0.049 & 0.048 \\
500 & -1.00 & 0.25 & -0.50 & 0.00 & 0.514 & 0.517 & 0.512 & 0.513 & 0.048 & 0.047 & 0.049 & 0.049 \\
500 & -1.00 & 0.25 & -0.50 & 0.25 & 0.514 & 0.514 & 0.508 & 0.506 & 0.047 & 0.047 & 0.049 & 0.051 \\
500 & -1.00 & 0.25 & -0.25 & 0.00 & 0.506 & 0.510 & 0.505 & 0.505 & 0.047 & 0.045 & 0.048 & 0.047 \\
500 & -1.00 & 0.25 & -0.25 & 0.25 & 0.507 & 0.511 & 0.502 & 0.503 & 0.046 & 0.046 & 0.050 & 0.047 \\
500 & 1.00 & 0.00 & -0.50 & 0.00 & 0.515 & 0.515 & 0.512 & 0.510 & 0.046 & 0.045 & 0.046 & 0.048 \\
500 & 1.00 & 0.00 & -0.50 & 0.25 & 0.516 & 0.519 & 0.515 & 0.515 & 0.047 & 0.044 & 0.048 & 0.047 \\
500 & 1.00 & 0.00 & -0.25 & 0.00 & 0.514 & 0.518 & 0.512 & 0.513 & 0.051 & 0.044 & 0.052 & 0.049 \\
500 & 1.00 & 0.00 & -0.25 & 0.25 & 0.512 & 0.516 & 0.512 & 0.513 & 0.047 & 0.047 & 0.047 & 0.048 \\
500 & 1.00 & 0.25 & -0.50 & 0.00 & 0.511 & 0.515 & 0.509 & 0.511 & 0.050 & 0.046 & 0.049 & 0.048 \\
500 & 1.00 & 0.25 & -0.50 & 0.25 & 0.511 & 0.512 & 0.512 & 0.511 & 0.051 & 0.049 & 0.052 & 0.050 \\
500 & 1.00 & 0.25 & -0.25 & 0.00 & 0.516 & 0.518 & 0.515 & 0.514 & 0.044 & 0.046 & 0.045 & 0.050 \\
500 & 1.00 & 0.25 & -0.25 & 0.25 & 0.517 & 0.521 & 0.517 & 0.518 & 0.047 & 0.044 & 0.047 & 0.046 \\
5,000 & -1.00 & 0.00 & -0.50 & 0.00 & 0.507 & 0.510 & 0.506 & 0.509 & 0.047 & 0.048 & 0.047 & 0.049 \\
5,000 & -1.00 & 0.00 & -0.50 & 0.25 & 0.504 & 0.504 & 0.505 & 0.505 & 0.050 & 0.049 & 0.049 & 0.048 \\
5,000 & -1.00 & 0.00 & -0.25 & 0.00 & 0.501 & 0.503 & 0.501 & 0.502 & 0.050 & 0.051 & 0.051 & 0.050 \\
5,000 & -1.00 & 0.00 & -0.25 & 0.25 & 0.505 & 0.501 & 0.505 & 0.502 & 0.046 & 0.046 & 0.046 & 0.046 \\
5,000 & -1.00 & 0.25 & -0.50 & 0.00 & 0.508 & 0.507 & 0.508 & 0.507 & 0.046 & 0.045 & 0.048 & 0.046 \\
5,000 & -1.00 & 0.25 & -0.50 & 0.25 & 0.508 & 0.508 & 0.506 & 0.508 & 0.049 & 0.051 & 0.047 & 0.050 \\
5,000 & -1.00 & 0.25 & -0.25 & 0.00 & 0.507 & 0.507 & 0.507 & 0.507 & 0.048 & 0.047 & 0.049 & 0.048 \\
5,000 & -1.00 & 0.25 & -0.25 & 0.25 & 0.504 & 0.504 & 0.505 & 0.505 & 0.050 & 0.047 & 0.048 & 0.048 \\
5,000 & 1.00 & 0.00 & -0.50 & 0.00 & 0.503 & 0.505 & 0.502 & 0.504 & 0.052 & 0.052 & 0.051 & 0.051 \\
5,000 & 1.00 & 0.00 & -0.50 & 0.25 & 0.505 & 0.507 & 0.506 & 0.508 & 0.046 & 0.045 & 0.046 & 0.044 \\
5,000 & 1.00 & 0.00 & -0.25 & 0.00 & 0.505 & 0.503 & 0.505 & 0.503 & 0.046 & 0.050 & 0.048 & 0.050 \\
5,000 & 1.00 & 0.00 & -0.25 & 0.25 & 0.502 & 0.504 & 0.503 & 0.505 & 0.050 & 0.046 & 0.049 & 0.046 \\
5,000 & 1.00 & 0.25 & -0.50 & 0.00 & 0.506 & 0.506 & 0.507 & 0.506 & 0.045 & 0.043 & 0.045 & 0.044 \\
5,000 & 1.00 & 0.25 & -0.50 & 0.25 & 0.505 & 0.506 & 0.508 & 0.509 & 0.051 & 0.049 & 0.050 & 0.050 \\
5,000 & 1.00 & 0.25 & -0.25 & 0.00 & 0.500 & 0.503 & 0.500 & 0.503 & 0.051 & 0.049 & 0.052 & 0.050 \\
5,000 & 1.00 & 0.25 & -0.25 & 0.25 & 0.505 & 0.503 & 0.507 & 0.506 & 0.049 & 0.048 & 0.050 & 0.047 \\ 
\hline
\end{tabular}
}
\vspace{0.3em}
\begin{minipage}{\linewidth}\footnotesize\raggedright $^*$ The Monte Carlo standard error is at most 0.003 for the mean p-value and 0.002 for the rejection rate. Scenario columns give the sample size $n$ and the coefficients of the reference model.\end{minipage}
\label{tab:h0_reject}
\end{table}
\end{landscape}

While our null hypothesis demands the moderate calibration of predicted
ITEs, the conditional approach, as discussed earlier, relies on the
condition that predicted baseline risks are calibrated within the strata
of predicted ITEs. The condition is satisfied in all above simulations.
To explore how the conditional test is affected when this condition is
violated, we extended the above simulations by distorting the predicted
baseline risks through a two-parameter transformation on the logit scale
(\(\mbox{logit}(\pi_0) = \beta_0 + \beta_1 \mbox{logit}(p_0)\)), in a
factorial design over sample size, \(\beta_0\), and \(\beta_1\). Results
indicate that both the BM and bridge conditional tests become
anti-conservative (with inflated type I error rate) - regardless of the
direction of miscalibration. This is an expected finding given that the
moments of the martingale (equations \mbox{\ref{condinc}} and
\mbox{\ref{condvar}}) depend on baseline risks, and thus miscalibrated
predicted risks can cause the path of the constructed random walk to
deviate from its expectation.

The second and third sets of simulations considered the power of these
tests under logit-linear (second set) and power (third set)
transformation of risks to create miscalibrated models. These two sets
consider various forms of miscalibration, including those where
predicted ITEs are systematically under- and over-estimates, as well as
where predicted ITEs are underestimates in some regions and
overestimates in others. These scenarios span both directions of model
misspecification: cases where the predicted model is simpler than the
data-generating truth (the truth being a non-linear or scaled departure
from the reference model), and cases where the predicted model is more
flexible than the truth (e.g., scenario s12, in which the model predicts
treatment-effect heterogeneity that is absent from the data-generating
process). The general design features modified versions of our previous
simulation setup for non-parametric tests of predicted
risks\cite{Sadatsafavi2024cumulcalib}. Here, we consider the reference
prediction model as having the following set of parameters:
\(\beta_0=0, \beta_x=0.25, \beta_a=-0.5, \beta_{xa}=0.25\). We generate
predictor values, treatment assignments, and predicted risks as in the
previous setup. Each scenario was run with \(n \in \{500,2500,10000\}\).

For the second set of simulations, we generate true outcome risks for
individuals under treatment as

\[\mbox{logit}(P(Y=1 | x,a)) = \beta_0 + \beta_x x + \alpha a + \gamma [\beta_a a + \beta_{xa} x a].\]

\begin{figure}[h!]
    \centering
    \caption{True calibration plots (actual risks on the Y axis and predicted risks on the X axis) for the second set of simulations}
    \includegraphics{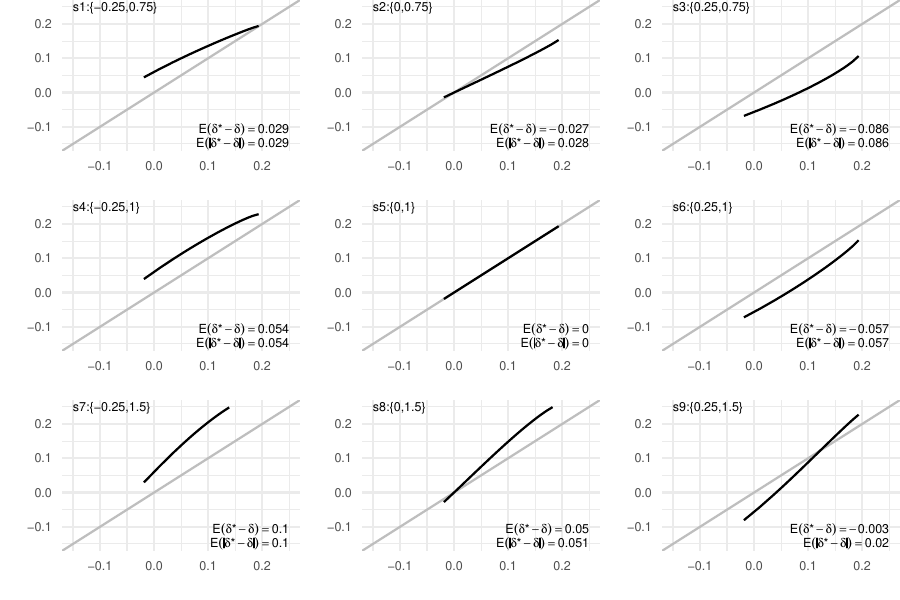}
    \label{fig:h1a_calibs_h}
\end{figure}

\begin{figure}[h!]
    \centering    
    \caption{Proportion of significant p-values (<0.05) for the second set of simulations}
    \includegraphics[width=\textwidth]{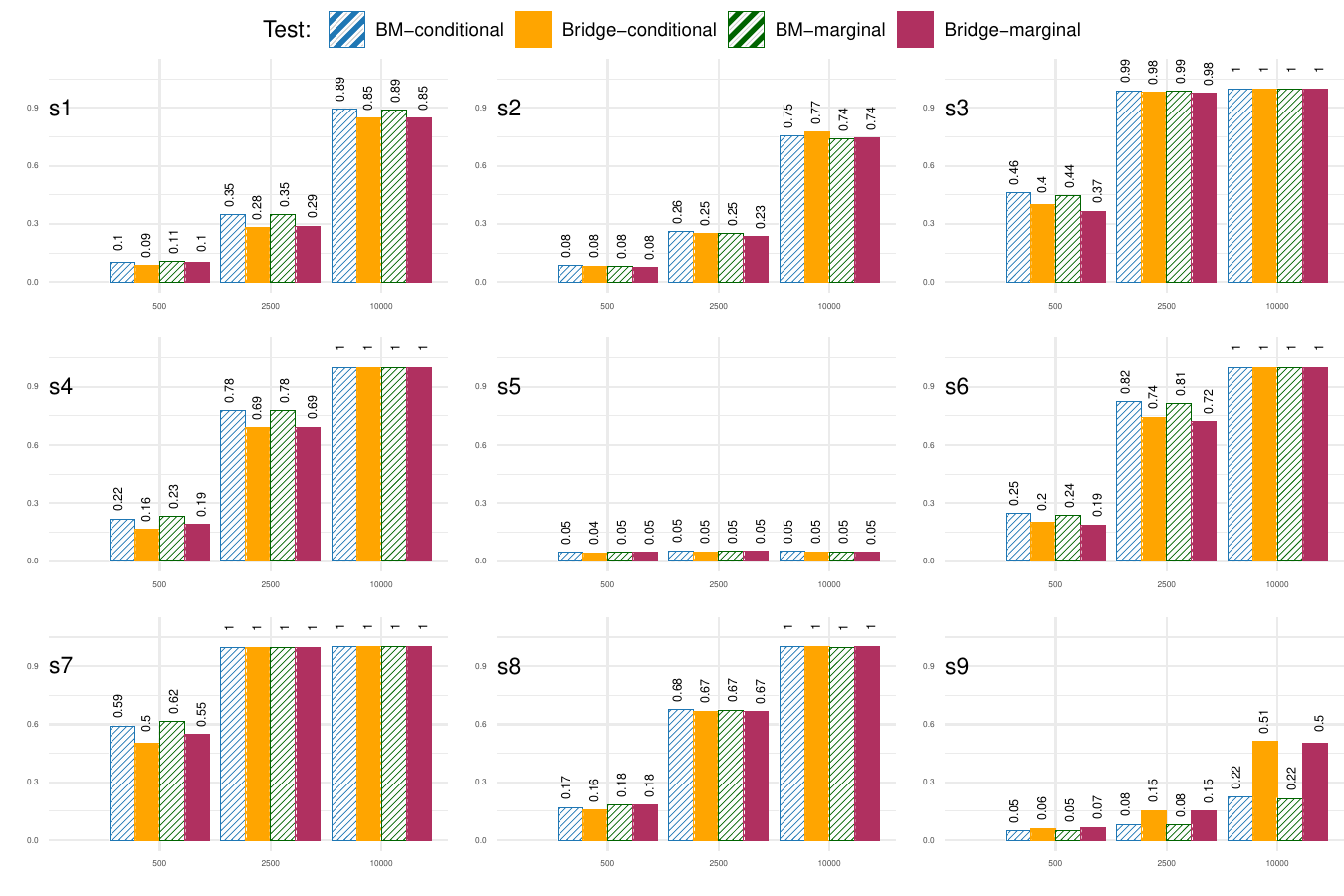}
    \label{fig:h1a_sim_bars}
    \begin{minipage}{\textwidth}\footnotesize\raggedright Each bar is based on 10,000 replications; the Monte Carlo standard error of each estimated rejection proportion is at most 0.005 (largest at a proportion of 0.5).\end{minipage}
\end{figure}

\noindent Note that the miscalibrated risks are linear on the logit
scale, with the two parameters \(\alpha\) and \(\gamma\) controlling,
respectively, the location and scale shifts of true risks among the
treated. The predicted and true risks among the control group remain
equal. The following parameter combinations for \(\{\alpha,\gamma\}\)
were examined: \{-0.25,0.75\}, \{0,0.75\}, \{0.25,0.75\}, \{-0.25,1\},
\{0,1\}, \{0.25,1\}, \{-0.25,1.5\}, \{0,1.5\}, \{0.25,1.5\}. The true
calibration plots are provided in Figure \ref{fig:h1a_calibs_h}. Two
summary measures are provided on each panel that quantify the extent of
miscalibration: the mean calibration error
(\(\mathbb{E}(\delta^*-\delta)\)) and the mean absolute calibration
error (\(\mathbb{E} | \delta^*-\delta|\)). The rejection probabilities
(at 5\% significance level) as a function of sample sizes are provided
in Figure \ref{fig:h1a_sim_bars}.

The third set of simulations considered non-linear forms of
miscalibration:

\[\mbox{logit}(P(Y=1 | x,a))=\alpha_a+\gamma_a\mbox{sign}(\mbox{logit}(\pi_a(x)))|\mbox{logit}(\pi_a(x))|^{\gamma_a}\].

\begin{figure}[h!]
    \centering
    \caption{True calibration plots (actual risks on the Y axis and predicted risks on the X axis) for the third set of simulations}
    \includegraphics{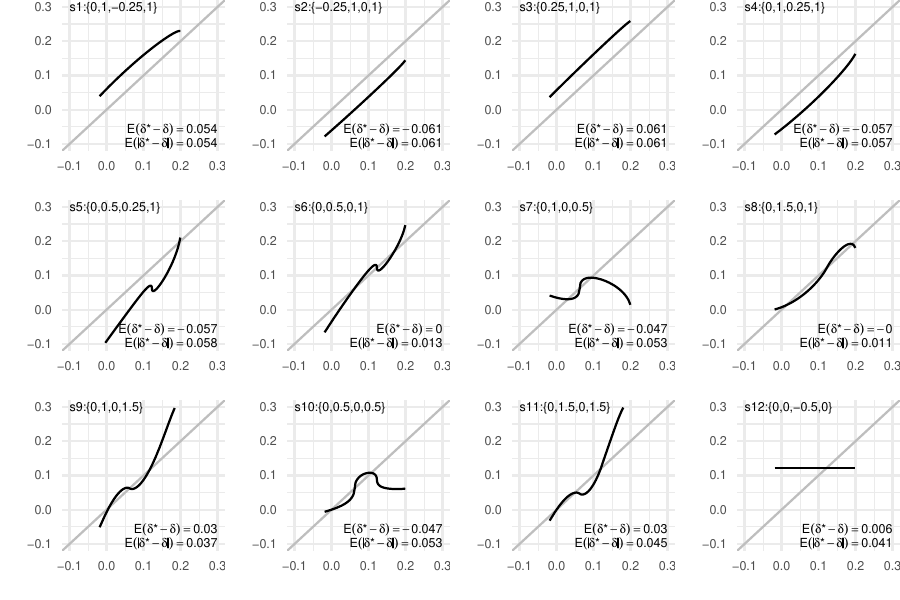}
    \label{fig:h1b_calibs_h}
\end{figure}

\begin{figure}[h!]
    \centering    
    \caption{Proportion of significant p-values (<0.05) for the third set of simulations}
    \includegraphics[width=\textwidth]{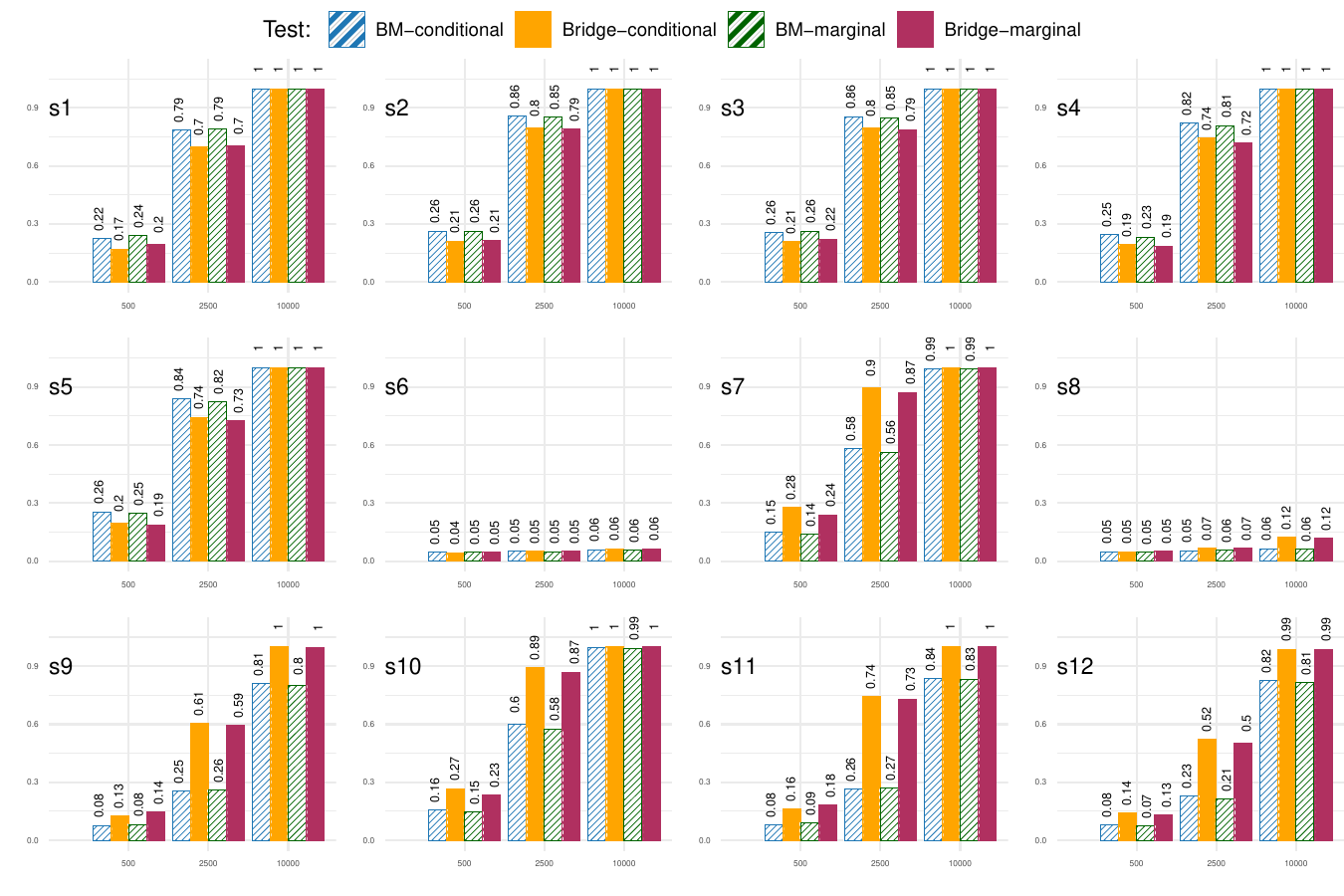}
    \label{fig:h1b_sim_bars}
    \begin{minipage}{\textwidth}\footnotesize\raggedright Each bar is based on 10,000 replications; the Monte Carlo standard error of each estimated rejection proportion is at most 0.005 (largest at a proportion of 0.5).\end{minipage}
\end{figure}

\noindent Here, \(\alpha_a\) introduces bias to the predicted risk while
the terms involving \(\gamma_a\) create an odd function that introduces
non-linear miscalibration. This function was applied to predicted risks
within each treatment arm, resulting in four degrees of freedom. The
following parameter combinations for
\(\{\alpha_0,\gamma_0,\alpha_1,\gamma_1\}\) were examined:
\{0,1,-0.25,1\}, \{-0.25,1,0,1\}, \{0.25,1,0,1\}, \{0,1,0.25,1\},
\{0,0.5,0.25,1\}, \{0,0.5,0,1\}, \{0,1,0,0.5\}, \{0,1.5,0,1\},
\{0,1,0,1.5\}, \{0,0.5,0,0.5\}, \{0,1.5,0,1.5\}, \{0,0,-0.5,0\}. In the
last scenario (s12), the ITE model falsely predicts treatment effect
heterogeneity. The true calibration plots for these scenarios are
provided in Figure \ref{fig:h1b_calibs_h}. Results of these simulations
are provided in Figure \ref{fig:h1b_sim_bars}.

Overall, across two sets of simulations, involving different forms and
degrees of miscalibrations, the tests demonstrated the expected
behavior: detecting miscalibrations with a power that generally
increases with the severity of miscalibration and the sample size. Some
other patterns are worth highlighting. The tests had low power for
detecting miscalibrated ITEs where the calibration function was close to
the identity line, even though the underlying predicted risks within
each arm were highly miscalibrated (e.g., scenarios s6 and s8 in the
second set). The conditional and marginal tests had nearly identical
power. This was not only the case in the second set (where baseline
predicted risks remained calibrated), but also in the third set, where
in many scenarios the baseline risks were miscalibrated. The two-part
bridge tests had generally non-inferior power compared with their
one-part BM counterparts, especially in scenarios where the
miscalibration function was highly non-linear (e.g., scenarios s7 and
s9-12 in the third set). The scenarios where the one-part tests had
marginally more power (s4, s6 in the second set, s1-s5 in the third set)
all pertain to cases where the calibration function is (almost) linear.
The bridge component of the two-part test is sensitive to non-linear
forms of miscalibration. As such, in these scenarios, this component is
generally non-significant, resulting in the p-value of Fisher's method
moving away from significance.

\section{Case study}\label{case-study}

We use data from the GUSTO-I study, which is a well-explored dataset in
predictive analytics methodological research\cite{GUSTO1993RCT}. GUSTO-I
was a multi-national clinical trial of streptokinase (SK) and tissue
plasminogen activator (TPA), two antithrombotic medications used for
immediate management of heart attacks. The primary endpoint was
mortality risk within 30 days. Given that death status could be verified
for all, there were no censored observations. The original study had
four study arms (SK+subcutaneous heparin, SK+intravenous heparin,
TPA+intravenous heparin, and SK+TPA+intravenous heparin). Combined
SK+TPA is not currently recommended due to high risk of bleeding, and
the original study ultimately combined the two SK groups (which showed
similar effects). As such, for this case study, we consider tPA alone
versus SK therapy, regardless of the mode of delivery of concomitant
heparin. We use this dataset to estimate ITEs for TPA versus SK therapy
after a heart attack. We assess the out-of-sample performance of the
resulting ITEs using the proposed visualizations, summary metrics, and
inference. All results are generated using the \emph{cumulcalib} R
package that accompanies this paper
(\url{https://github.com/resplab/cumulcalib}), with exemplary code
provided in Appendix 3.

Similar to previous case studies, we use the non-US subset of the sample
to develop the ITE models, and use the US sub-sample to evaluate it. As
this case study is an illustration, we will develop two ITE models. The
large-development-sample model uses the entire development sample
(\(n=\) 13,342), while the small-development-sample model is based on a
random subset consisting of 10\% of the development sample. Similarly,
we use two validation samples. The large validation sample uses the
entire validation sample (\(n=\) 17,168), while the small validation
sample uses a random subset consisting of 10\% of the validation sample.
Outcome risks (the probability of dying in the first 30 days after the
event) are as follows: 0.074 for the large development sample, 0.069 for
the small development sample, 0.067 for the large validation sample, and
0.061 for the small validation sample. The corresponding ATEs were
0.007, -0.017, 0.012, and 0.024.

We fit models of the form
\(\mbox{logit}P(Y=1)=\beta_0+\beta_1[\text{female}] + \beta_2[\text{age}] + \beta_3[\text{AMI location - other}] + \beta_4[\text{AMI location anterior}] + \beta_5[\text{previous AMI}] + \beta_6 [\text{Killip score}] + \beta_7 [min(\text{systolic blood pressure}, 100)] + \beta_8[\text{pulse rate}] + \beta_9 [\text{a}] + \beta_{10}[\text{a:female}] + \beta_{11}[\text{a:age}]\),
with \(a\) indicating treatment and \(a:b\) indicating the first-order
interaction between two predictors. Note that the treatment indicator
(TPA:1, SK:0) enters the model both as a main effect and with
first-order interaction with sex and age (reflecting our hypothesis that
these variables are treatment effect modifiers on the logit scale). The
coefficients of these models are provided in the footnotes of the
cumulative calibration plots below.

\subsection{Large-development-sample
model}\label{large-development-sample-model}

\begin{figure}[h]
    \centering
    \caption{Visualization of the \{S\} process for the large-development-sample model, an example whose cumulative calibration plot shows no systematic departure from calibration. Top row: large validation sample; bottom row: small validation sample. Left column: conditional approach; right column: marginal approach.}
    \begin{tabular}{cc}
        \includegraphics[width=0.45\textwidth]{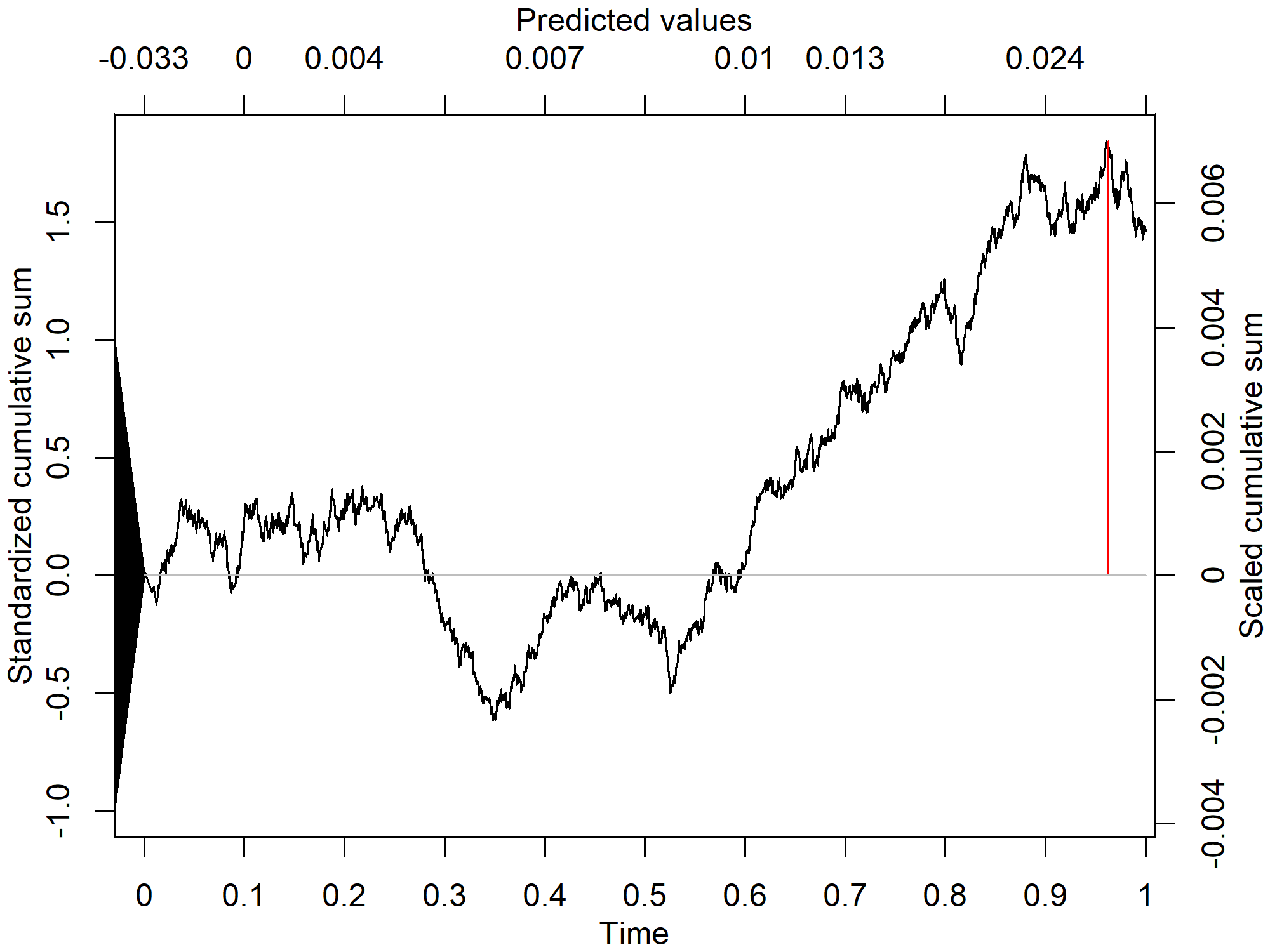} &
        \includegraphics[width=0.45\textwidth]{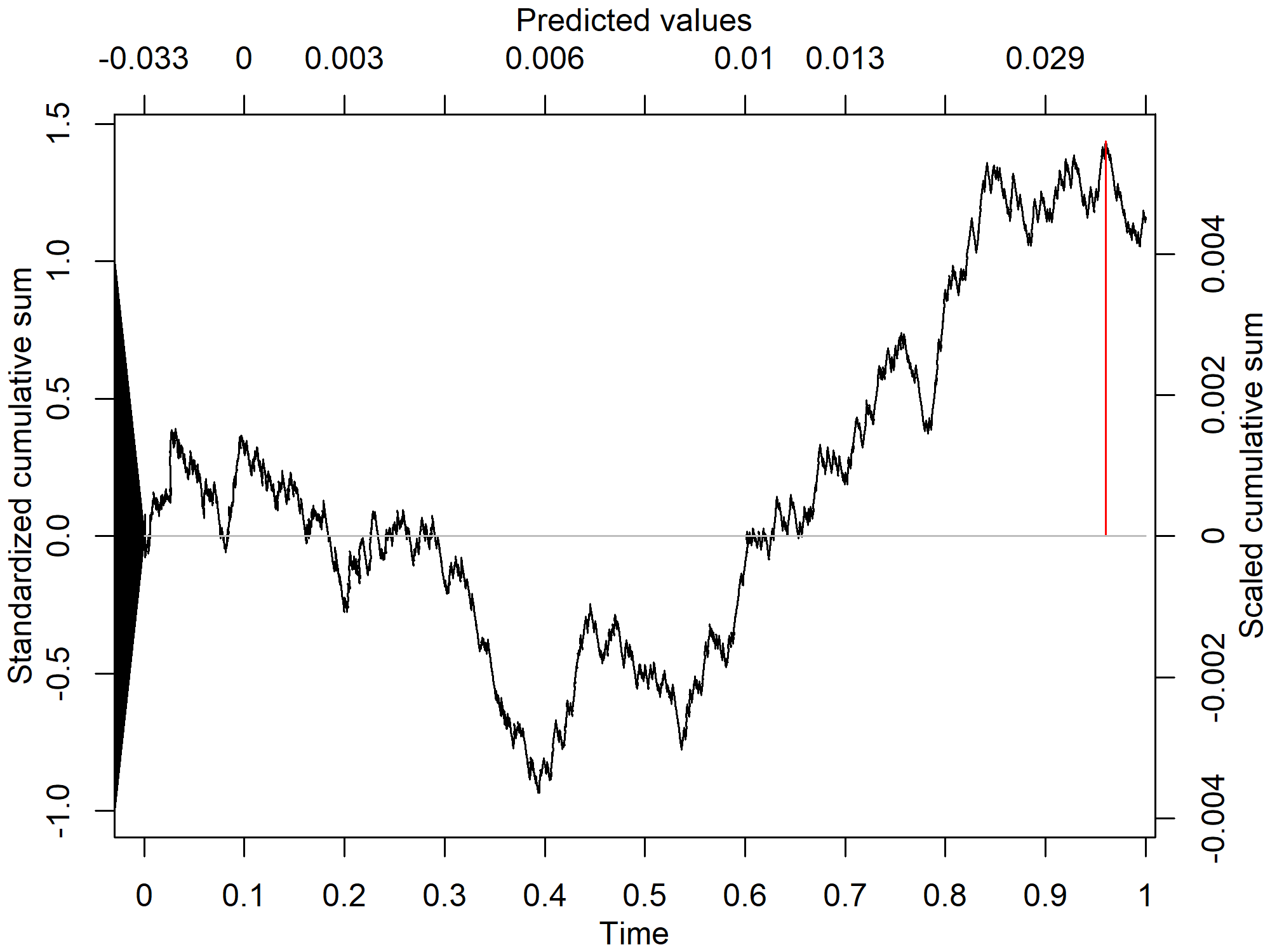} \\
        \includegraphics[width=0.45\textwidth]{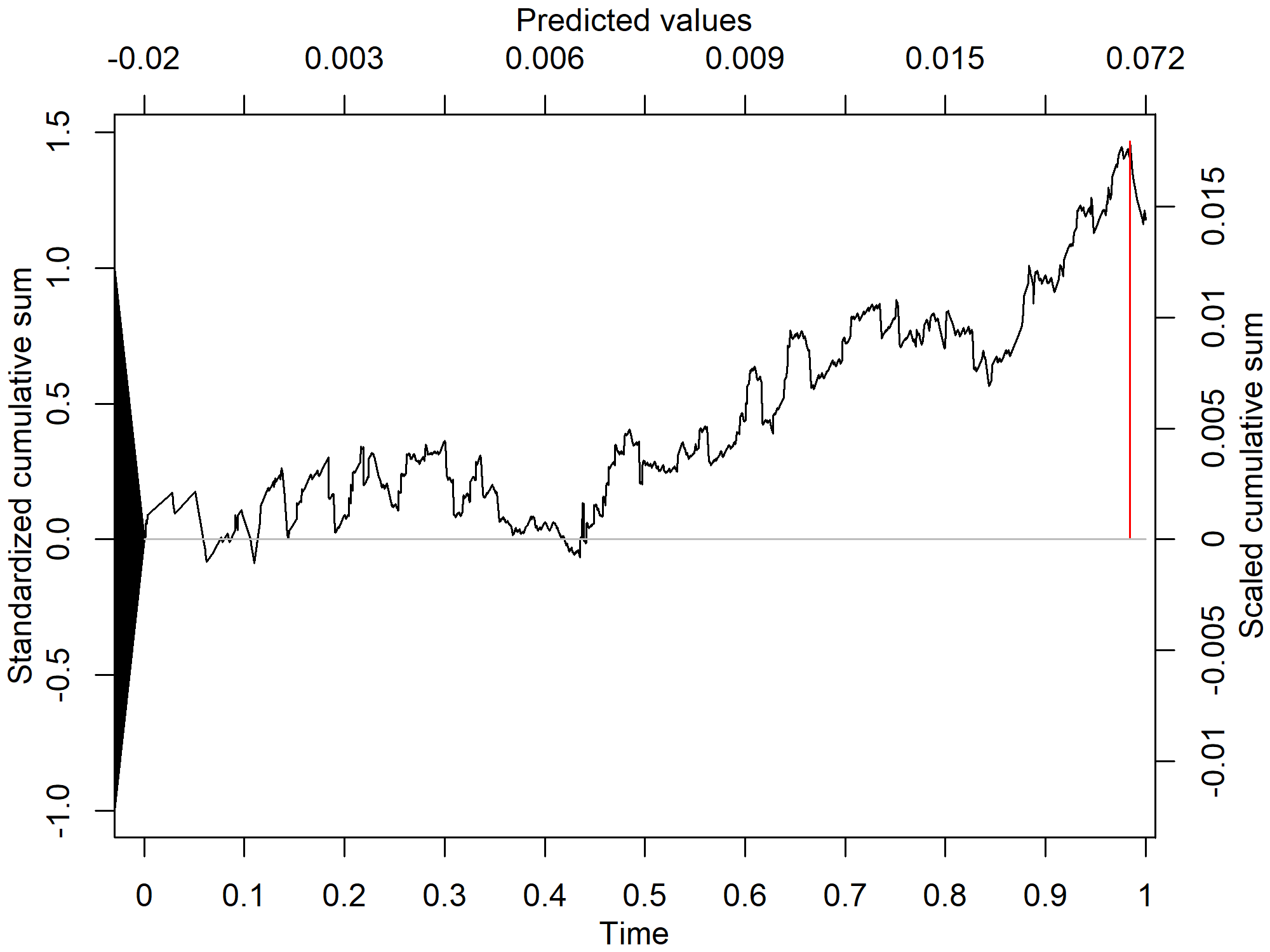} &
        \includegraphics[width=0.45\textwidth]{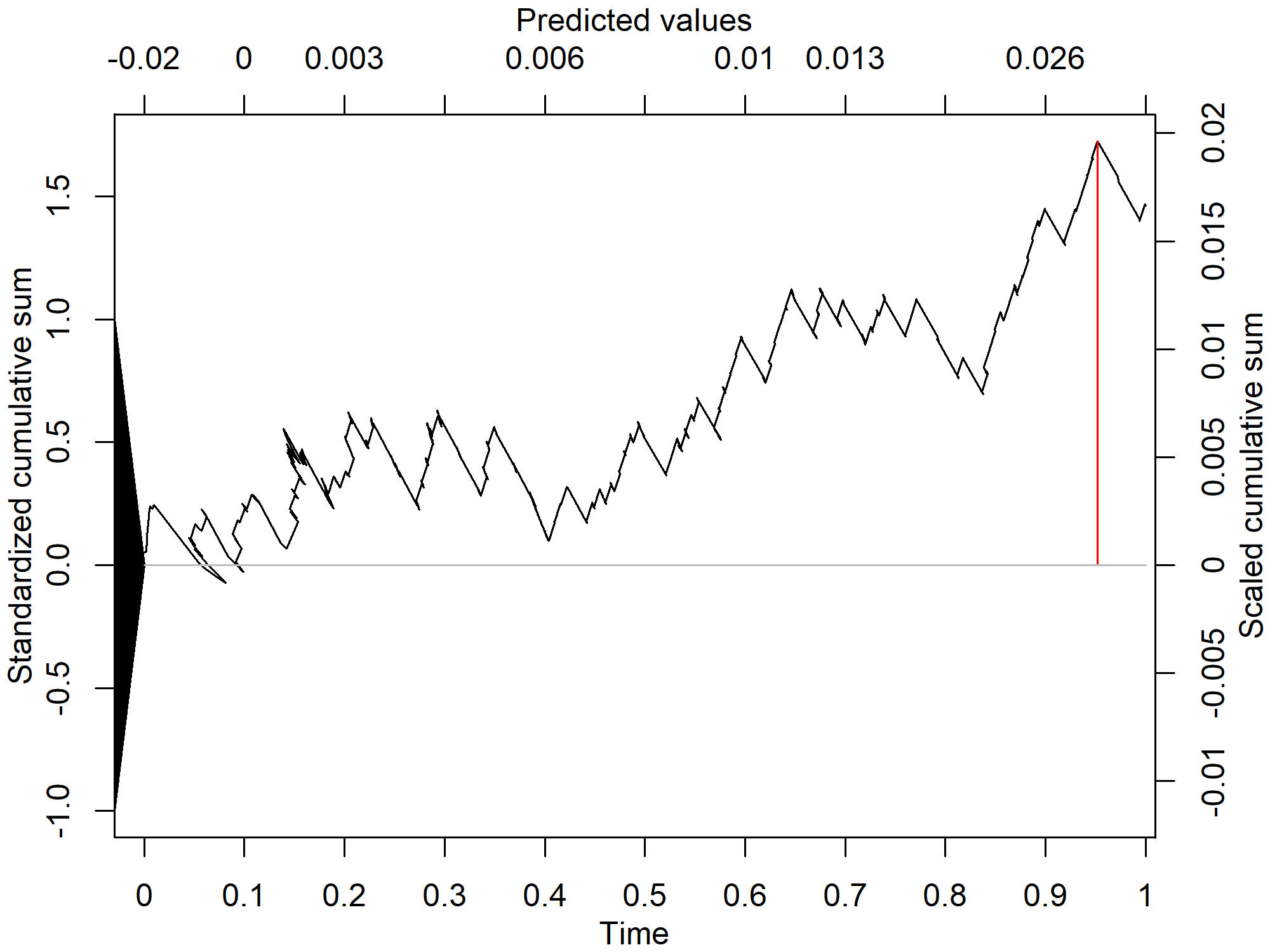} \\
    \end{tabular}
    \label{fig:gusto_modelL}
    \vspace{0.5em}
    \begin{minipage}{\textwidth}
    \footnotesize\raggedright
    \textit{How to read these plots:} The primary (bottom and left) axes show the time and location components of the standardized \{S\} process (Equation \ref{SProcess}). Under moderate calibration this path behaves like a standard Brownian motion, fluctuating randomly around zero with no systematic drift, whereas a systematic drift or a large excursion indicates miscalibration. The secondary (top) axis shows the predicted ITE ($\delta$), which maps one-to-one to time; the secondary (right) axis shows the scaled, un-standardized process, whose terminal value is $C_n$ (the mean calibration error). The solid vertical line marks the location and direction of the maximum absolute cumulative calibration error, $C^*$. Model coefficients ($\beta_0$--$\beta_{11}$) are, respectively: -1.2246, 0.3147, 0.0684, 0.3926, 0.5353, 0.4596, 0.7375, -0.0790, 0.0179, -0.8846, -0.0405, 0.0111.
    \par
    \end{minipage}
\end{figure}

Figure \ref{fig:gusto_modelL} presents the visualization of the \{S\}
process for this model for the large (top row) and small (bottom row)
validation samples, for the conditional (left column) and marginal
(right column) approaches. Details of this visualization are similar to
those proposed for the risk prediction
context\cite{Sadatsafavi2024cumulcalib}. The primary X- and Y-axes
pertain to, respectively, time and location components of the
standardized (\{S\}) process. The secondary (top) X-axis is utilized to
show the predicted ITEs (\(\delta\)s), noting that there is a 1:1
mapping between a given time value and a predicted value. The secondary
(right) Y-axis shows the scaled (but not standardized) process; the
terminal value of the random walk on this axis corresponds to \(C_n\),
providing a convenient visual assessment of mean calibration error. The
solid vertical line marks the location and direction of the maximum
absolute cumulative calibration error (\(C^*\)), the point at which the
standardized process is farthest from zero. Note the occasional negative
jumps in the marginal graph for the small validation sample
(bottom-right panel).

The values of \(C_n\) and \(C^*\), and the p-value for the null
hypothesis that the model is moderately calibrated, are reported in
Table \ref{tab:gusto_modelL}. For example, in the large validation
sample under the conditional approach, \(C_n=0.0055\) and \(C^*=0.0070\)
(on the risk-reduction scale). Overall, the figures and summary indices
are compatible with this ITE model being calibrated: the random walk
does not drift systematically from its expected path. p-values indicate
that the data are not strongly against the claim that the model is
calibrated in either sample.

\begin{table}[ht]
\centering
\caption{Large-development-sample model validation results. $C_n$ is the (signed) mean calibration error and $C^*$ the (non-negative) maximum absolute cumulative calibration error, both on the risk-reduction scale. The p-value is that of the two-part bridge test (the combination, via Fisher's method, of its mean-calibration and bridge components).}
\resizebox{\textwidth}{!}{%
\begin{tabular}{llccc}
\hline
Validation sample & Approach & \begin{tabular}{@{}c@{}} $C_n$ \\ (mean calibration error) \end{tabular} & \begin{tabular}{@{}c@{}} $C^*$ \\ (maximum absolute cumulative error) \end{tabular} & \begin{tabular}{@{}c@{}} p-value \\ (bridge test) \end{tabular} \\
\hline
Large & Conditional & 0.0055 & 0.0070 & 0.0633 \\
Large & Marginal & 0.0045 & 0.0056 & 0.0554 \\
\hline
Small & Conditional & 0.0144 & 0.0179 & 0.5432 \\
Small & Marginal & 0.0166 & 0.0196 & 0.4054 \\
\hline
\end{tabular}
} 
\label{tab:gusto_modelL}
\end{table}

\subsection{Small-development-sample
model}\label{small-development-sample-model}

\begin{figure}[h]
    \centering
    \caption{Visualization of the \{S\} process for the small-development-sample model, an example of a clearly miscalibrated ITE model. Rows and columns are as in Figure \ref{fig:gusto_modelL}.}
    \begin{tabular}{cc}
        \includegraphics[width=0.45\textwidth]{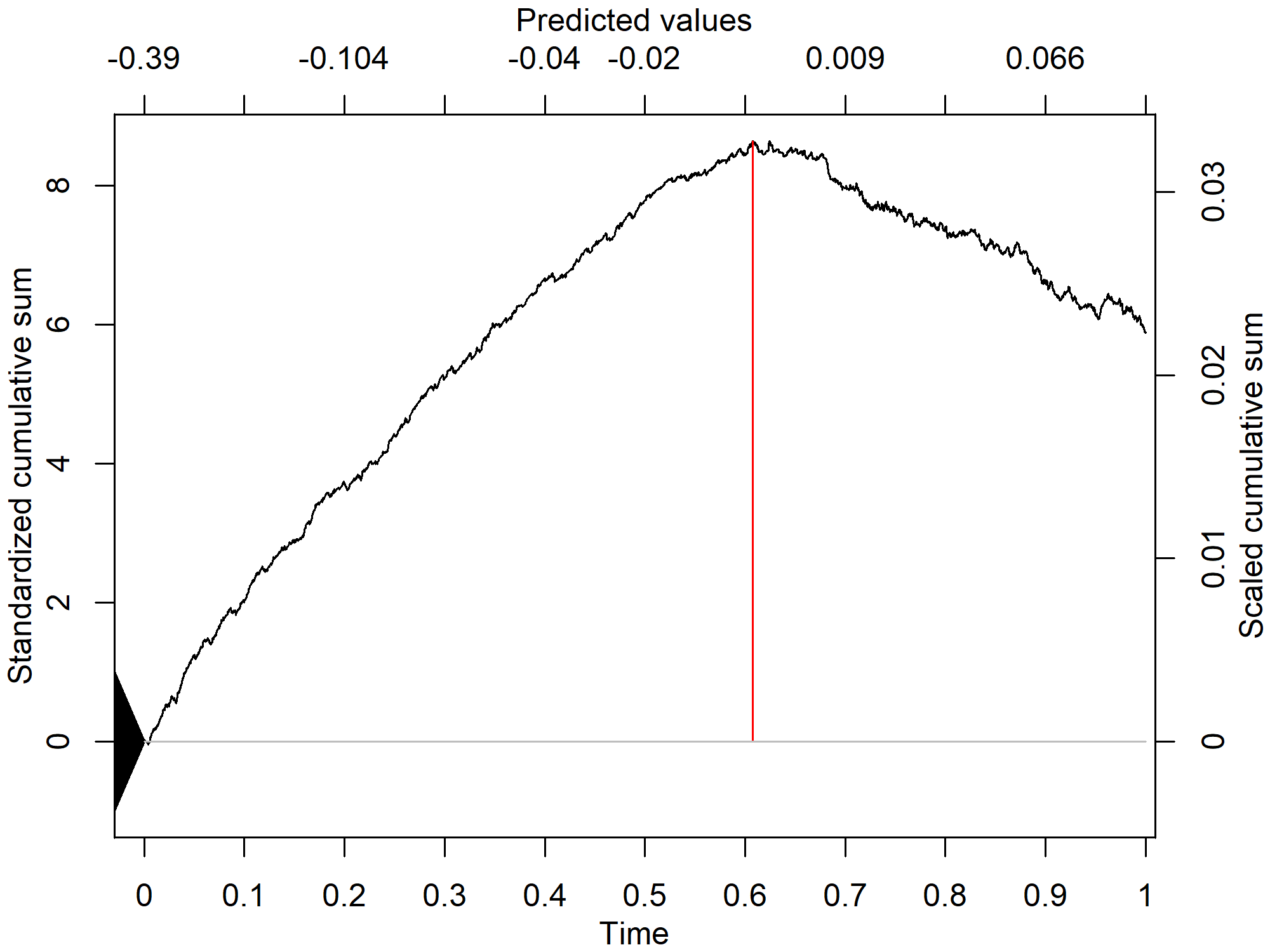} &
        \includegraphics[width=0.45\textwidth]{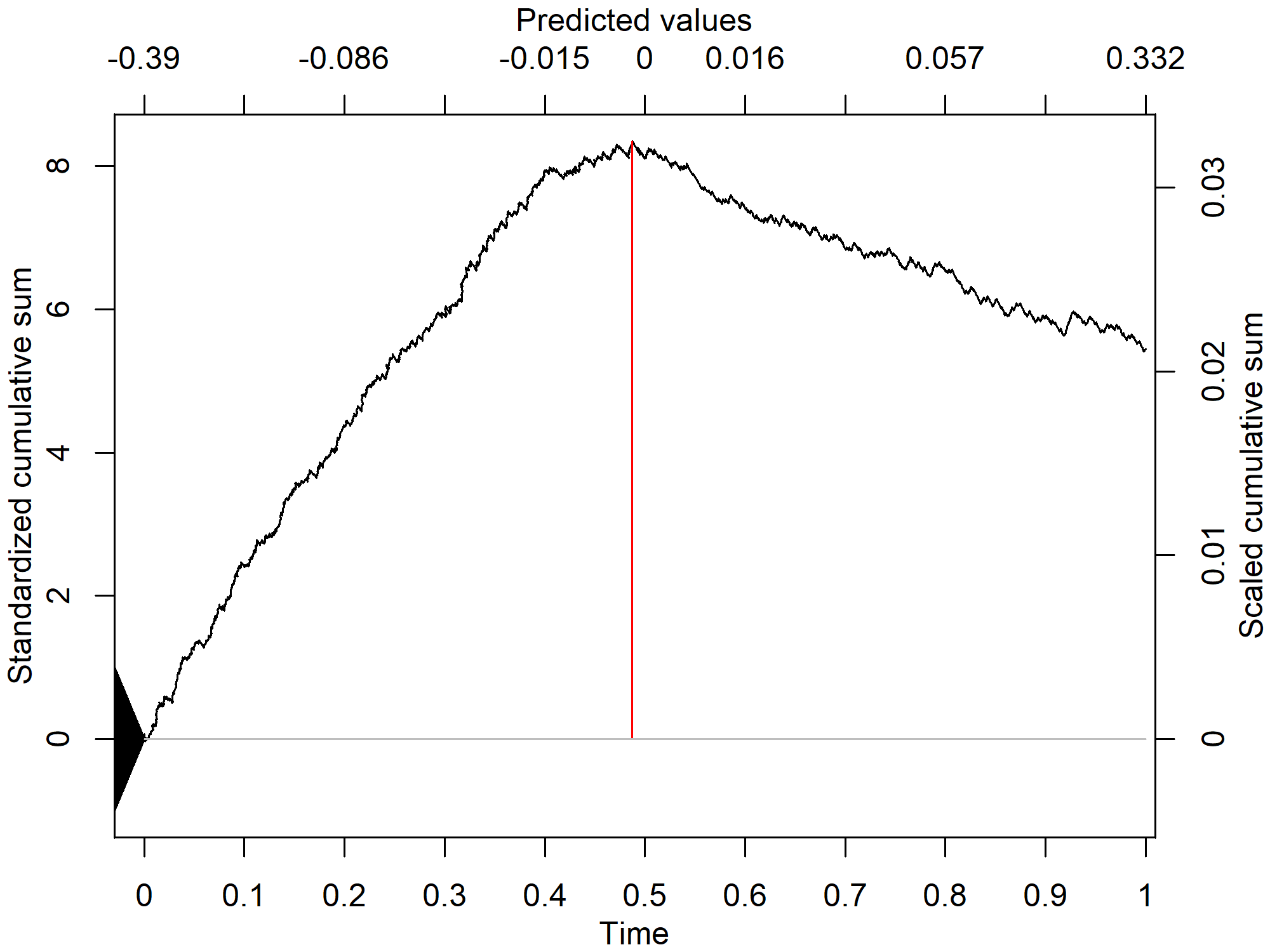} \\
        \includegraphics[width=0.45\textwidth]{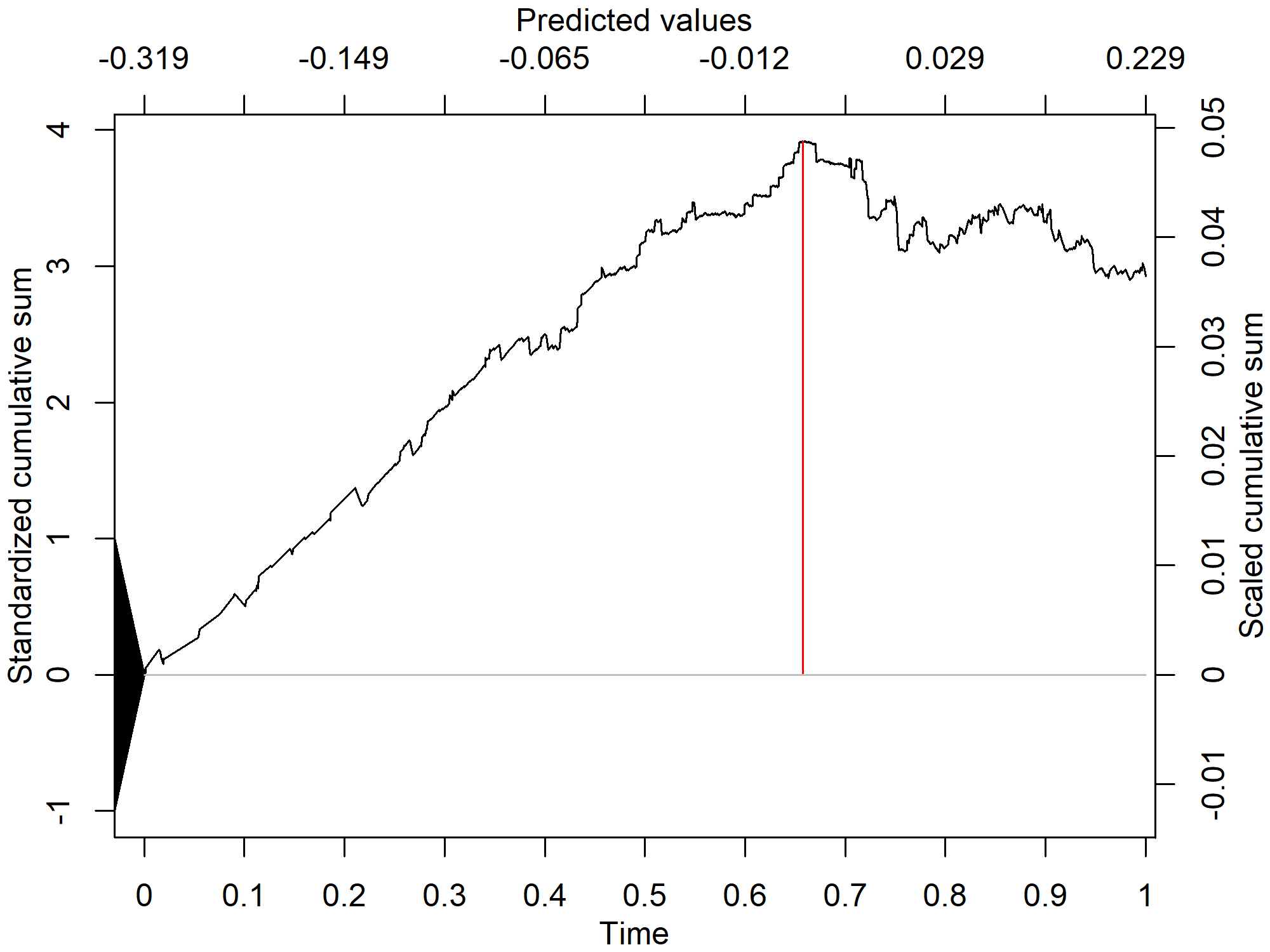} &
        \includegraphics[width=0.45\textwidth]{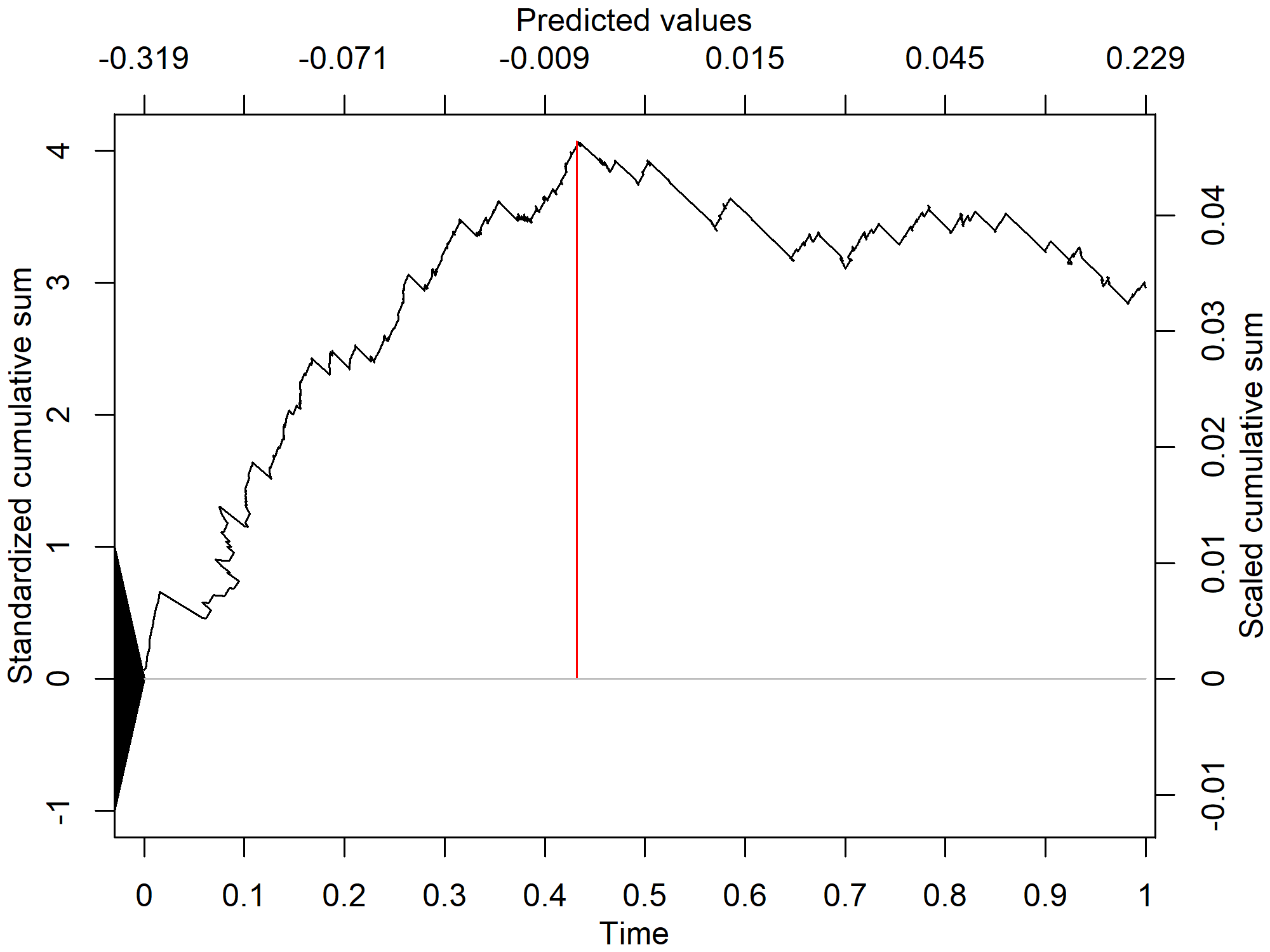} \\
    \end{tabular}
    \vspace{0.5em}
    \begin{minipage}{\textwidth}
    \footnotesize\raggedright
    Axes and plot elements are as described in Figure \ref{fig:gusto_modelL}. Here the standardized path rises and then falls (an inverted U): prediction errors accumulate positively for small predicted ITEs (benefit under-estimated) and negatively for large ones (benefit over-estimated) - the signature of an over-fitted model that exaggerates the spread of predicted ITEs. The large $C^*$ and small bridge-test p-values (Table \ref{tab:gusto_modelS}) confirm strong evidence of miscalibration. Model coefficients ($\beta_0$--$\beta_{11}$) are, respectively: -2.2635, 0.4721, 0.0529, -0.5550, 0.4923, 0.4515, 0.6840, -0.0622, 0.0210, -2.3426, -1.4945, 0.0435.
    \par
    \end{minipage}
    \label{fig:gusto_modelS}
\end{figure}

Graphical and numerical results are provided in Figure
\ref{fig:gusto_modelS} and Table \ref{tab:gusto_modelS}, respectively.
Here, the model is clearly miscalibrated in both validation samples. The
prediction errors accumulate on the left side of the graph (up to around
a predicted ITE of 0, as indicated by the secondary X-axis), indicating
that the model under-estimates the actual benefit when the predicted ITE
is small (actual ITE \textgreater{} predicted ITE, resulting in positive
errors). The curve then declines, indicating that the model
over-estimates the benefit for larger predicted ITEs (actual ITE
\textless{} predicted ITE, resulting in negative errors). As noted
earlier, this is a signature of an over-fitted model that makes
exaggerated predictions: it over-predicts benefit for patients with high
predicted ITE and under-predicts it for those with low predicted ITE.
Correspondingly, the summary indices are large, indicating a substantial
degree of miscalibration, and the bridge p-values are small, indicating
strong evidence for miscalibration. In such instances, the location of
\(C^*\) provides useful auxiliary information: taking the conditional
approach in the large validation sample as an example, the maximum
absolute cumulative deviation occurs at a predicted ITE of -0.0034.
Predicted ITEs around this value are approximately calibrated, whereas
those below it are downwardly biased (benefit under-estimated) and those
above it are upwardly biased (benefit over-estimated).

\begin{table}[ht]
\centering
\caption{Small-development-sample model validation results. Columns are as in Table \ref{tab:gusto_modelL}: $C_n$ is the (signed) mean calibration error and $C^*$ the (non-negative) maximum absolute cumulative calibration error (both on the risk-reduction scale), and the p-value is that of the two-part bridge test.}
\resizebox{\textwidth}{!}{%
\begin{tabular}{llccc}
\hline
Validation sample & Approach & \begin{tabular}{@{}c@{}} $C_n$ \\ (mean calibration error) \end{tabular} & \begin{tabular}{@{}c@{}} $C^*$ \\ (maximum absolute cumulative error) \end{tabular} & \begin{tabular}{@{}c@{}} p-value \\ (bridge test) \end{tabular} \\
\hline
Large & Conditional & 0.0223 & 0.0328 & $<$0.0001 \\
Large & Marginal    & 0.0212 & 0.0325 & $<$0.0001 \\
\hline
Small & Conditional & 0.0364 & 0.0488 & $<$0.0001 \\
Small & Marginal    & 0.0337 & 0.0464 & $<$0.0001 \\
\hline
\end{tabular}
} 
\label{tab:gusto_modelS}
\end{table}

\section{Discussion}\label{discussion}

In this work we proposed novel non-parametric methods for the assessment
of moderate calibration of predicted ITEs for binary outcomes. These
methods are applicable when a pre-specified ITE model is being validated
in a new sample. We have demonstrated that moderate calibration can be
assessed without resorting to regularization if calibration is examined
on the cumulative sum domain. We proposed an estimator for cumulative
ITE errors and showed that a properly scaled version of this estimator
will result in a stochastic process that, under the null hypothesis that
the ITE model is moderately calibrated, converges point-wise to the
standard Brownian motion. We showed that the conditional moments of this
estimator depend on the true outcome risks. For ITE models that predict
risk, such predictions can be used under the condition that they are
calibrated, giving rise to our proposed `conditional' approach. However,
the investigator might not be willing to assume those risks are
calibrated; indeed, not all ITE models generate predicted risks. For
these instances, we proposed an approximate `marginal' approach that
solely relies on predicted ITEs. These developments enabled us to extend
the previously proposed visualization, summarizing indices, and tests
for moderate calibration of predicted risk to the predicted ITE domain.
In a case study we laid out our suggestions for the visualization of
cumulative prediction errors and the interpretation of proposed tests.
The proposed tests are implemented in the \emph{cumulcalib} R package
with exemplary code provided in Appendix 3.

For inference, our proof-of-concept simulations confirmed the desired
properties of the resulting tests. Simulations under the null
hypothesis, where ITE predictions were moderately calibrated, showed
that these asymptotic tests generally maintain their type I error rate,
even with sample sizes of 500, which we consider modest for an ITE model
validation study. In simulations where the null hypothesis was violated,
the power of these tests increased with larger sample sizes and with
more severe miscalibrations, albeit the power could be low if the true
calibration function was close to the identity line. The two-part
(bridge) test had generally more power than the one-part test (at times
significantly), an unsurprising finding given that this test offers two
independent opportunities for detecting miscalibration. Because it rests
on between-arm differences, assessing ITE calibration is more demanding,
in terms of power, than assessing risk calibration. This is related to
the notion that detecting treatment-by-covariate interaction requires
significantly larger sample sizes\cite{Brookes2004interactionPower}. In
our simulations the tests were well-powered at \(n=10,000\) and detected
pronounced miscalibration at \(n=2,500\), but had limited power at
\(n=500\). In modest samples, therefore, high p-values, especially in
the presence of a large \(C^*\), should not be interpreted as evidence
in favor of calibration; rather, that the sample does not offer strong
evidence against the claim that the model is calibrated. We generally
advise against significance testing at an arbitrary level for model
validation. Rather, p-values are better interpreted as continuous
measures of the compatibility between the data and the hypothesis of
moderate calibration, complementing, not replacing, the reported
magnitude of miscalibration\cite{RafiGreenland2020pval}.

How can these developments be used in practice? Some models generate
ITEs from underlying predicted risks under alternative treatment
decisions. Such predicted risks might be used for decision making in
tandem with predicted ITEs. Correspondingly, one might demand that the
model generate calibrated predictions across all its outputs. Requiring
an ITE model to be calibrated on both predicted risk and ITE domains is
equal to requiring it to be calibrated on predicted risks within each
treatment group. One option is therefore to assess calibration
separately for risks within each treatment group. However, this does not
put the focus on calibration of ITEs, which might be of independent
interest. Indeed, miscalibration of risks within each group might cancel
out and result in calibrated ITEs; reciprocally, otherwise subtle
miscalibrations of risks can augment and result in more pronounced
miscalibration of ITEs. These patterns might not be obvious when
predicted risks are assessed independently. Therefore, even for ITE
models that also predict risk we propose an assessment of the
calibration of ITEs, especially the plot of the \{S\} process. If the
predicted risks are deemed calibrated (or the model is modified to
correct for miscalibration), one can then use the conditional approach,
for which a functional CLT could be proven. However, the marginal
approach also seems to have desirable properties while not requiring the
assumption of calibrated predicted risks. The latter remains the only
option for models that generate predicted ITEs without modeling risks.

We propose visualization of the standardized cumulative calibration
plot, as well as reporting \(C_n\), i.e., mean calibration error, and
\(C^*\), i.e., maximum absolute cumulative prediction error. If
inference is desired, we suggest the use of the two-part `bridge' test
over the one-part BM test (all these suggestions are exemplified in our
case study). Investigators often assess mean calibration (by comparing
the predicted and observed ATEs) before moderate calibration. As mean
calibration is a necessary condition for moderate calibration, such
sequential inference alters the power of assessing moderate calibration.
The bridge test provides an opportunity for joint inference that
preserves the desired overall type I error rate while reducing type II
error rate because of its higher power. Further, in many instances, if
mean calibration assessment is not satisfactory, one might first decide
to address the miscalibrated ATEs before embarking on moderate
calibration assessment. For example, for an ITE model that predicts
risks under control and intervention strategies using a logit link
function, one might fix a biased ATE prediction by adjusting the
intercept of the logit function separately within each treatment arm.
This approach results in average predicted risks exactly matching the
average observed risks within treatment groups, in turn leading to an
exact match between the predicted and observed ATEs. In these instances,
the one-part BM test is not a valid inference method, as the terminal
value of the \{S\} process is guaranteed to be zero. In such instances,
one can apply the bridge component of the bridge test in isolation to
assess moderate calibration conditional on mean calibration having been
achieved.

There are several areas for future research. As a first step, we
proposed these approaches for binary outcomes and when the sample is
from a randomized experiment. An important extension, with wide
applicability, would be for other outcome types, in particular survival
outcomes. For such outcomes, the challenge is that the observed response
is not only a function of predicted risk, but also the distribution of
censoring events and the shape of the hazard function. As for extending
this approach to non-randomized samples, the challenge is that \(a\),
\(\delta\), and \(Y\) can be dependent in such settings, violating the
martingale properties underpinning the proposed stochastic processes. A
potential path forward is the application of
inverse-probability-of-treatment weighting schemes that can result in
pseudo-populations where treatment assignment can be considered random.
However, the application of a functional CLT when observations are
weighted would require additional technical developments. Further, other
known properties of the Brownian motion can be used to propose
alternative tests. In addition to the test based on maximum absolute
deviation, Arrieta-Ibarra et al proposed a test based on the maximum
range of the Brownian motion
(\(\max(W)-\min(W)\))\cite{ArrietaIbarra2022BM}. In the same vein, a
test based on the maximum range of the Brownian bridge will result in a
new two-part test. Another potential test would be based on the square
of the area under the Brownian motion and Brownian bridge. The resulting
tests might have more power as they capture the behavior of the \{S\}
process across its entire path. Lastly, our simulation studies were
aimed at investigating the expected properties of the proposed tests.
The power of these tests in detecting specific types of miscalibrations
should be investigated in dedicated simulation studies. This should
include a deeper investigation of the consequences of predicted risks
not being calibrated for the conditional approach, and investigating the
marginal approach in more varied settings given its heuristic
underpinnings. Relatedly, while our framework conditions on the model's
predicted ITEs and is thus agnostic to how they were generated (and the
proposed marginal approach does not require arm-specific predicted risks
to be available), our simulations produced predicted ITEs as differences
of arm-specific predicted risks. Assessing predictions from estimators
that target the treatment effect directly (e.g., causal forests or
R-learners) is a further useful direction for future work.

A further extension of this framework is to study calibration of
predictions across the domain of another variable. Conventionally, as in
this work, we assess calibration as the closeness of predicted and
actual quantities across the strata of predicted quantities (ITEs in our
case). However, this closeness can also be investigated across the
strata of another variable. For example, it is often reasonable to
investigate whether a risk prediction or an ITE model makes calibrated
predictions among individuals of different age. This involves assessing
the closeness of predicted and observed risks or ITEs within strata
defined by age. The core idea of non-parametric calibration assessment
is to transport the conditional mean
\(\mathbb{E}(Y^{(0)}-Y^{(1)}-\delta|\delta=z)\) into the cumulative
domain \(\mathbb{E}(Y^{(0)}-Y^{(1)}-\delta|\delta<z)\). The premise is
that the latter quantity can be estimated, with proper scaling, from the
cumulative sums of prediction errors, once ordering the sample from the
small to large values of \(\delta\). Imagine we are now interested in
assessing the calibration of predictions across the strata of a
continuous variable \(h\) (e.g., age). The quantity of interest
therefore becomes \(\mathbb{E}(Y^{(0)}-Y^{(1)}-\delta|h=z)\), which
again can be cast onto the cumulative sum domain
\(\mathbb{E}(Y^{(0)}-Y^{(1)}-\delta|h<z)\). The latter can be estimated
via scaled cumulative sums, this time after ordering the observations
(\(\delta, a, Y\)) from the small to large values of \(h\). The
subsequent steps will remain as described in this work, including
visualization and inference. This form of calibration assessment is
especially relevant for ITE models that combine predicted risks with an
estimate of treatment effect. As an example, the ACCEPT model, a risk
prediction model for exacerbations in patients with obstructive lung
disease\cite{Safari2022ACCEPT2}, is also used to predict the ITE (risk
reduction) of treatment with azithromycin. This is achieved by applying
a constant treatment effect, estimated from relevant RCTs, to predicted
risks. It might be of interest to check if the assumption of a constant
effect holds across the strata of predicted risks. For this purpose,
once the observations in the sample are ordered from small to large
values of predicted risks, the \{S\} process can be constructed as
before.

Presently, most prediction models focus on outcome risk under a single
control treatment (e.g., standard of care). Even though these models are
not designed to individualize treatment decisions, they are often used
as such, typically by assuming that individuals at high outcome risk
should be prioritized for active treatment. On the other hand, by
predicting individualized treatment effect, ITE models provide a
rational framework for medical decision-making. Perhaps due to
recognizing the importance of ITE models, and the increasing
availability of data to support their development, there is a rising
interest in both the methodology and application of such models.
Calibration is an often overlooked aspect of prediction models in
general\cite{VanCalster2019AchillesHeel}, but is arguably even more
critical for predicted ITEs due to their direct relevance in informing
treatment decisions. Therefore, it seems timely to highlight the
importance of calibration for ITE models and promote objective,
reproducible methods for its assessment.

\clearpage

\bibliographystyle{unsrtnat}
\bibliography{references}

\clearpage

\section*{\texorpdfstring{Appendix 1: Equations for increments and their
conditional expectations for
\(B_k\)}{Appendix 1: Equations for increments and their conditional expectations for B\_k}}\label{appendix-1-equations-for-increments-and-their-conditional-expectations-for-b_k}
\addcontentsline{toc}{section}{Appendix 1: Equations for increments and
their conditional expectations for \(B_k\)}

\begin{equation*}
 D_k = B_k-B_{k-1}=\left\{
\begin{matrix}
k\frac{\sum_{i=1}^{k}{(1-a_i)Y_i}}{\sum_{i=1}^{k}(1-a_i)}-(k-1)\frac{\sum_{i=1}^{k-1}{(1-a_i)Y_i}}{\sum_{i=1}^{k-1}(1-a_i)}-\frac{\sum_{i=1}^{k-1}{a_iY_i}}{\sum_{i=1}^{k-1} a_i} & & a_k=0 \\

\frac{\sum_{i=1}^{k-1}{(1-a_i)Y_i}}{\sum_{i=1}^{k-1}(1-a_i)} -  \left[k\frac{\sum_{i=1}^{k}{ta_iY_i}}{\sum_{i=1}^{k}a_i}-(k-1)\frac{\sum_{i=1}^{k-1}{a_iY_i}}{\sum_{i=1}^{k-1}a_i}\right] & & a_k=1
\end{matrix}
\right.
\end{equation*}

\begin{equation*}
  \mu_k =E(B_k-B_{k-1}|\mathcal{F}_{k-1})=\left\{
\begin{matrix}
k\frac{\sum_{i=1}^{k-1}{(1-a_i)Y_i}+\pi_k^*}{\sum_{i=1}^{k}(1-a_i)}-(k-1)\frac{\sum_{i=1}^{k-1}{(1-a_i)Y_i}}{\sum_{i=1}^{k-1}(1-a_i)}-\frac{\sum_{i=1}^{k-1}{a_iY_i}}{\sum_{i=1}^{k-1} a_i} & a_k=0 \\

\frac{\sum_{i=1}^{k-1}{(1-a_i)Y_i}}{\sum_{i=1}^{k-1}(1-a_i)} -  \left[k\frac{\sum_{i=1}^{k-1}{a_iY_i}+(\pi_k^*-\delta_k^*)}{\sum_{i=1}^{k}a_i}-(k-1)\frac{\sum_{i=1}^{k-1}{a_iY_i}}{\sum_{i=1}^{k-1}a_i}\right] & a_k=1 \\
\end{matrix}
\right.,
\end{equation*}

where \(\pi^*_k=\mathbb{E}(Y|\delta=\delta_k, a=0)\) and
\(\delta^*_k=\mathbb{E}(Y|\delta=\delta_k, a=0)-\mathbb{E}(Y|\delta=\delta_k, a=1)\).

\clearpage

\section*{Appendix 2: Proof of the Lindeberg
condition}\label{appendix-2-proof-of-the-lindeberg-condition}
\addcontentsline{toc}{section}{Appendix 2: Proof of the Lindeberg
condition}

We need to show that

\[ \lim_{n \to \infty} \frac{1}{s_n^{2}}\sum_{k=1}^{n}\left[\mathbb{E}(D_k-\mu_k)^{2}\,\mathbf 1_{\{|D_k-\mu_k| \,>\, \epsilon s_n\}}\right] = 0\]

for all \(\epsilon \,>\, 0\), where

\[D_k - \mu_k = k \left( (1 - a_k) \frac{Y_k - \pi^*_k}{\sum_{i=1}^{k}(1 - a_i)} - a_k \frac{Y_k - \pi^*_k + \delta^*_k}{\sum_{i=1}^{k} a_i} \right), \]
and

\[s^2_n=\sum_{k=1}^n k^2\left[(1-a_k)\frac{\pi^*_k(1-\pi^*_k)}{(\sum_{i=1}^{k}(1-a_i))^2}+a_k\frac{(\pi^*_k-\delta^*_k)(1-\pi^*_k+\delta^*_k)}{(\sum_{i=1}^{k}a_i)^2}\right].\]

Our proof is for when individuals are allocated to the treatment or
control groups by a random mechanism that is unrelated to the response
and the predicted benefits. Let \(A_k = \sum_{i=1}^{k} a_i\). The
following regularity conditions are required:
\(\lim_{n \to \infty} A_n/n=\lambda\), where \(0<\lambda<1\); and the
true conditional risks \(\pi^*_k\) and \(\pi^*_k-\delta^*_k\) are
bounded away from 0 and 1, ensuring
\(\lim_{n \to \infty} s_n = \infty\).

Without loss of generality we assume \(0 < \lambda \le 0.5\). Because
the numerators of the fractions in the formula for \(D_k-\mu_k\) are
always between \(-1\) and \(1\), we have

\[|D_k - \mu_k| < k \left( (1 - a_k) \frac{1}{k - A_k} + a_k \frac{1}{A_k} \right) \le \frac{k}{\min(k - A_k, A_k)}.\]

\noindent Note that the denominator in the non-zero term above counts
the subjects, among the first \(k\), in the same arm as the \(k\)th
subject, and is therefore at least one; hence \(|D_k-\mu_k| < k\). So,
defining

\[Z_k=\frac{k}{\min(k-A_k,A_k)},\]

\noindent we have

\[\mathbb{E}\left[(D_k-\mu_k)^{2} \,\mathbf 1_{\{|D_k-\mu_k| > \epsilon s_n\}}\right] \le \mathbb{E}\left[(D_k-\mu_k)^{2} \,\mathbf 1_{\{Z_k > \epsilon s_n\}}\right].\]

\noindent Therefore, it we establish the limit with the term on the
right-hand side instead, the goal is achieved. Further, as the term
\(\epsilon s_n\) grows without bound, if we change it to any fixed value
and prove the result, it will be sufficient. Thus, we show that

\[\lim_{n \to \infty} \frac{1}{s_n^{2}}\sum_{k=1}^{n} \mathbb{E}\left[(D_k-\mu_k)^{2}\,\mathbf 1_{\{Z_k> \lambda^{-1}+\xi\}}\right] = 0\]

\noindent for some fixed \(\xi > 0\) (note that
\(\lim_{k\to\infty} \frac{k}{\min(k-A_k,\,A_k)}=\lambda^{-1})\). But

\[
\begin{aligned}
P\left(Z_k>\lambda^{-1}+\xi\right) &= P\left(\frac{k}{\min{\left(k-A_k,A_k\right)}}>\lambda^{-1}+\xi\right) \\
 &= P\left(\min{\left(k-A_k,A_k\right)}<\frac{k}{\lambda^{-1}+\xi}\right) \\
 &\le P \left(\min{\left(2\lambda k-A_k,A_k\right)}<\frac{k}{\lambda^{-1}+\xi}\right) \\
 &= P \left( \min(\lambda k - A_k, A_k - \lambda k) < \frac{k}{\lambda^{-1} + \xi} - \lambda k \right) = P\left(|A_k - \lambda k| > k\frac{\lambda^2\xi}{1+\lambda\xi}\right).
\end{aligned}
\]

Due to the assumed independence between the treatment allocations and
the response and predicted benefits, we can assume ordering the sample
on predicted benefits (as part of constructing the \(\{S\}\) process) is
equivalent to a random permutation of the treatment allocations. As
such, by Hoeffding's inequality for sampling without replacement
(Hoeffding, 1963, Section 6),

\[ \forall c>0: P\!\bigl(\left|A_k-k\tfrac{A_n}{n}\right| > kc\bigr) \le 2e^{-2kc^2}.\]

\noindent Writing \(c_\xi=\frac{\lambda^2\xi}{1+\lambda\xi}\), and
noting that \(A_n/n \to \lambda\) implies
\(|A_n/n-\lambda| \le c_\xi/2\) for all sufficiently large \(n\), we
obtain, for all such \(n\) and all \(k \le n\),

\[ P\!\bigl(|A_k-\lambda k| > kc_\xi\bigr) \le P\!\bigl(\left|A_k-k\tfrac{A_n}{n}\right| > kc_\xi/2\bigr) \le 2e^{-kc_\xi^2/2}.\]

\noindent Since \(|D_k-\mu_k| < k\),

\[\sum_{k=1}^{n}\mathbb{E}\left[(D_k-\mu_k)^2\,\mathbf 1_{\{Z_k> \lambda^{-1}+\xi\}}\right] \le \sum_{k=1}^{n} k^2\, P\left(Z_k>\lambda^{-1}+\xi\right) \le \sum_{k=1}^{\infty} 2k^2e^{-kc_\xi^2/2} < \infty,\]

\noindent where the last series converges as the exponential factor
dominates the polynomial term (ratio test). As \(s_n\) grows without
bound,

\[ \lim_{n\to\infty} \frac{1}{s_n^{2}}\sum_{k=1}^{n}\mathbb{E}\left[(D_k-\mu_k)^{2}\mathbf 1_{\{Z_k> \lambda^{-1}+\xi\}}\right] = 0.\]

\bigskip

\noindent\textbf{QED.}

\bigskip
\clearpage

\clearpage

\section*{Appendix 3: Exemplary R
code}\label{appendix-3-exemplary-r-code}
\addcontentsline{toc}{section}{Appendix 3: Exemplary R code}

This code uses the \emph{cumulcalib} R package and the GUSTO data as an
example of the implementation of the proposed tests. To illustrate a
model with clearly detectable miscalibration, we deliberately develop
the ITE model on a small random subset (one tenth) of the development
data --- an over-fitted model, mirroring the small-development-sample
model of the case study --- and validate it on a one-tenth random subset
of the validation sample.

The GUSTO data are available from the \emph{predtools} package (on
CRAN). The official version of the \emph{cumulcalib} package can also be
installed from CRAN; its development version is available on github. The
output below was generated with version 0.1.0 of the package, archived
under the tag \texttt{v0.1.0-SiMBenefit}.

\begin{Shaded}
\begin{Highlighting}[]
\DocumentationTok{\#\#\#Un{-}comment and run the lines below to install the exact version of the}
\DocumentationTok{\#\#\#cumulcalib package used for this paper (tag v0.1.0{-}SiMBenefit)}
\CommentTok{\#library(remotes)}
\CommentTok{\#remotes::install\_github("resplab/cumulcalib@v0.1.0{-}SiMBenefit", force=TRUE)}

\FunctionTok{library}\NormalTok{(predtools) }\CommentTok{\#For GUSTO data}
\FunctionTok{data}\NormalTok{(}\StringTok{"gusto"}\NormalTok{)}
\FunctionTok{library}\NormalTok{(cumulcalib)}

\NormalTok{df }\OtherTok{\textless{}{-}}\NormalTok{ gusto[}\SpecialCharTok{{-}}\FunctionTok{which}\NormalTok{(gusto}\SpecialCharTok{$}\NormalTok{tx}\SpecialCharTok{==}\StringTok{"SK+tPA"}\NormalTok{),] }\CommentTok{\# Removing the double therapy arm}
\NormalTok{df}\SpecialCharTok{$}\NormalTok{a }\OtherTok{\textless{}{-}} \FunctionTok{ifelse}\NormalTok{(df}\SpecialCharTok{$}\NormalTok{tx}\SpecialCharTok{==}\StringTok{"tPA"}\NormalTok{, }\DecValTok{1}\NormalTok{, }\DecValTok{0}\NormalTok{)}\SpecialCharTok{*}\DecValTok{1} \CommentTok{\# 1 for TPA 0 for other (SKA or SKA combo)}
\NormalTok{df}\SpecialCharTok{$}\NormalTok{Y }\OtherTok{\textless{}{-}}\NormalTok{ df}\SpecialCharTok{$}\NormalTok{day30}
\NormalTok{df}\SpecialCharTok{$}\NormalTok{kill }\OtherTok{\textless{}{-}}\NormalTok{ (}\FunctionTok{as.numeric}\NormalTok{(df}\SpecialCharTok{$}\NormalTok{Killip)}\SpecialCharTok{\textgreater{}}\DecValTok{1}\NormalTok{)}\SpecialCharTok{*}\DecValTok{1} \DocumentationTok{\#\# Killip Class Binary (0 = KC 1)}

\CommentTok{\# Development sample: non{-}US; validation sample: US (as in the case study)}
\NormalTok{data\_dev }\OtherTok{\textless{}{-}}\NormalTok{ df[}\SpecialCharTok{!}\NormalTok{df}\SpecialCharTok{$}\NormalTok{regl }\SpecialCharTok{\%in\%} \FunctionTok{c}\NormalTok{(}\DecValTok{1}\NormalTok{, }\DecValTok{7}\NormalTok{, }\DecValTok{9}\NormalTok{, }\DecValTok{10}\NormalTok{, }\DecValTok{11}\NormalTok{, }\DecValTok{12}\NormalTok{, }\DecValTok{14}\NormalTok{, }\DecValTok{15}\NormalTok{),] }\CommentTok{\# Non{-}US data}
\NormalTok{data\_val }\OtherTok{\textless{}{-}}\NormalTok{ df[df}\SpecialCharTok{$}\NormalTok{regl }\SpecialCharTok{\%in\%} \FunctionTok{c}\NormalTok{(}\DecValTok{1}\NormalTok{, }\DecValTok{7}\NormalTok{, }\DecValTok{9}\NormalTok{, }\DecValTok{10}\NormalTok{, }\DecValTok{11}\NormalTok{, }\DecValTok{12}\NormalTok{, }\DecValTok{14}\NormalTok{, }\DecValTok{15}\NormalTok{),]  }\CommentTok{\# US data}

\CommentTok{\# Deliberately develop on a small (over{-}fitted) subset and validate on a subset,}
\CommentTok{\# to obtain an ITE model with clearly detectable miscalibration}
\FunctionTok{set.seed}\NormalTok{(}\DecValTok{122}\NormalTok{)}
\NormalTok{df\_dev }\OtherTok{\textless{}{-}}\NormalTok{ data\_dev[}\FunctionTok{sample}\NormalTok{(}\FunctionTok{nrow}\NormalTok{(data\_dev), }\FunctionTok{round}\NormalTok{(}\FunctionTok{nrow}\NormalTok{(data\_dev) }\SpecialCharTok{/} \DecValTok{10}\NormalTok{)), ]}
\NormalTok{df\_val }\OtherTok{\textless{}{-}}\NormalTok{ data\_val[}\FunctionTok{sample}\NormalTok{(}\FunctionTok{nrow}\NormalTok{(data\_val), }\FunctionTok{round}\NormalTok{(}\FunctionTok{nrow}\NormalTok{(data\_val) }\SpecialCharTok{/} \DecValTok{10}\NormalTok{)), ]}

\NormalTok{ite\_model }\OtherTok{\textless{}{-}} \FunctionTok{glm}\NormalTok{(Y }\SpecialCharTok{\textasciitilde{}}\NormalTok{ sex }\SpecialCharTok{+}\NormalTok{ age }\SpecialCharTok{+}\NormalTok{ miloc }\SpecialCharTok{+}\NormalTok{ pmi }\SpecialCharTok{+}\NormalTok{ kill }\SpecialCharTok{+} \FunctionTok{pmin}\NormalTok{(sysbp,}\DecValTok{100}\NormalTok{) }\SpecialCharTok{+}\NormalTok{ pulse }\SpecialCharTok{+}
\NormalTok{                   a }\SpecialCharTok{+}\NormalTok{ a}\SpecialCharTok{:}\NormalTok{sex }\SpecialCharTok{+}\NormalTok{ a}\SpecialCharTok{:}\NormalTok{age, }\AttributeTok{data =}\NormalTok{ df\_dev,}
                  \AttributeTok{family=}\FunctionTok{binomial}\NormalTok{(}\AttributeTok{link=}\StringTok{"logit"}\NormalTok{))}

\CommentTok{\#Creating counterfactual predicitons and predicted ite (h)}
\NormalTok{df\_ }\OtherTok{\textless{}{-}}\NormalTok{ df\_val}
\NormalTok{df\_}\SpecialCharTok{$}\NormalTok{a }\OtherTok{\textless{}{-}} \DecValTok{0}
\NormalTok{p0 }\OtherTok{\textless{}{-}} \FunctionTok{predict}\NormalTok{(ite\_model, }\AttributeTok{newdata=}\NormalTok{df\_, }\AttributeTok{type=}\StringTok{"response"}\NormalTok{)}
\NormalTok{df\_}\SpecialCharTok{$}\NormalTok{a }\OtherTok{\textless{}{-}} \DecValTok{1}
\NormalTok{p1 }\OtherTok{\textless{}{-}} \FunctionTok{predict}\NormalTok{(ite\_model, }\AttributeTok{newdata=}\NormalTok{df\_, }\AttributeTok{type=}\StringTok{"response"}\NormalTok{)}

\NormalTok{df\_val}\SpecialCharTok{$}\NormalTok{p0 }\OtherTok{\textless{}{-}}\NormalTok{ p0}
\NormalTok{df\_val}\SpecialCharTok{$}\NormalTok{h }\OtherTok{\textless{}{-}}\NormalTok{ p0}\SpecialCharTok{{-}}\NormalTok{p1}
\end{Highlighting}
\end{Shaded}

\begingroup
\fontsize{8pt}{10pt}\selectfont

\begin{Shaded}
\begin{Highlighting}[]
\CommentTok{\#Approach 1, Brownian motion test}
\NormalTok{res1 }\OtherTok{\textless{}{-}} \FunctionTok{cumulcalibITE}\NormalTok{(df\_val}\SpecialCharTok{$}\NormalTok{Y, df\_val}\SpecialCharTok{$}\NormalTok{h, df\_val}\SpecialCharTok{$}\NormalTok{a, df\_val}\SpecialCharTok{$}\NormalTok{p0, }\AttributeTok{method =} \StringTok{"BM"}\NormalTok{)}
\FunctionTok{summary}\NormalTok{(res1)}
\end{Highlighting}
\end{Shaded}

\begin{verbatim}
## Moderate calibration assessment of individualized treatment effects
## Approach: conditional
## C_n (mean calibration error): 0.0363985965193455
## C* (maximum cumulative calibration error): 0.0488039015226009  (observed benefit > predicted)
##   Location of maximum cumulative error: time = 0.657662473334137, predicted ITE = -0.00331168848044092
##   Shape: the cumulative error reverses around predicted ITE = -0.00331 (observed benefit exceeds predicted below this point; observed benefit falls below predicted above it)
## Method: One-part Brownian motion (BM)
## S* (test statistic for cumulative calibration error): 3.91889501088661
## p-value: 0.000177911699480449
\end{verbatim}

\begin{Shaded}
\begin{Highlighting}[]
\CommentTok{\# lines="vh" draws the vertical test{-}statistic line AND the horizontal}
\CommentTok{\# significance threshold; the package default ("v") omits the threshold}
\FunctionTok{plot}\NormalTok{(res1, }\AttributeTok{stats\_config =} \FunctionTok{list}\NormalTok{(}\AttributeTok{lines =} \StringTok{"vh"}\NormalTok{))}
\end{Highlighting}
\end{Shaded}

\pandocbounded{\includegraphics[keepaspectratio]{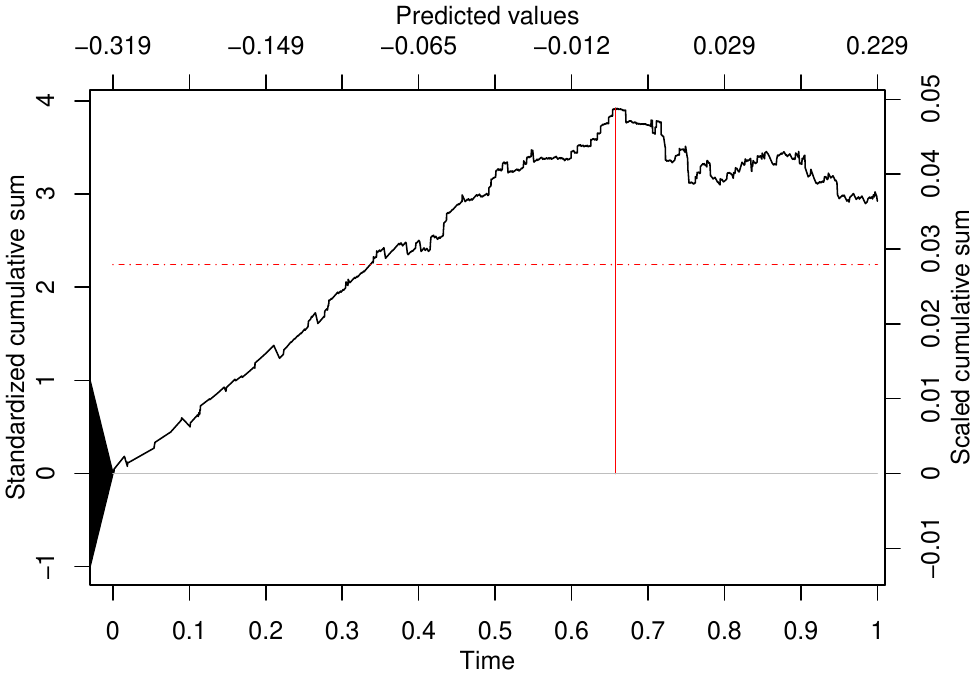}}

\begin{Shaded}
\begin{Highlighting}[]
\CommentTok{\#Approach 2, Brownian bridge test}
\NormalTok{res2 }\OtherTok{\textless{}{-}} \FunctionTok{cumulcalibITE}\NormalTok{(df\_val}\SpecialCharTok{$}\NormalTok{Y, df\_val}\SpecialCharTok{$}\NormalTok{h, df\_val}\SpecialCharTok{$}\NormalTok{a, df\_val}\SpecialCharTok{$}\NormalTok{p0, }\AttributeTok{method =} \StringTok{"BB"}\NormalTok{)}
\FunctionTok{summary}\NormalTok{(res2)}
\end{Highlighting}
\end{Shaded}

\begin{verbatim}
## Moderate calibration assessment of individualized treatment effects
## Approach: conditional
## C_n (mean calibration error): 0.0363985965193455
## C* (maximum cumulative calibration error): 0.0488039015226009  (observed benefit > predicted)
##   Location of maximum cumulative error: time = 0.657662473334137, predicted ITE = -0.00331168848044092
##   Shape: the cumulative error reverses around predicted ITE = -0.00331 (observed benefit exceeds predicted below this point; observed benefit falls below predicted above it)
## Method: Two-part Brownian bridge (BB)
## S_n (Z score for mean calibration error): 2.92276383347919
## B* (test statistic for maximum absolute bridged calibration error): 1.99735740841779
## Component-wise p-values: mean calibration=0.00346939553914561 | Distance (bridged)=0.000685250527216574
## Combined p-value (Fisher's method): 3.3163615011933e-05
\end{verbatim}

\begin{Shaded}
\begin{Highlighting}[]
\FunctionTok{plot}\NormalTok{(res2, }\AttributeTok{stats\_config =} \FunctionTok{list}\NormalTok{(}\AttributeTok{lines =} \StringTok{"vh"}\NormalTok{))}
\end{Highlighting}
\end{Shaded}

\pandocbounded{\includegraphics[keepaspectratio]{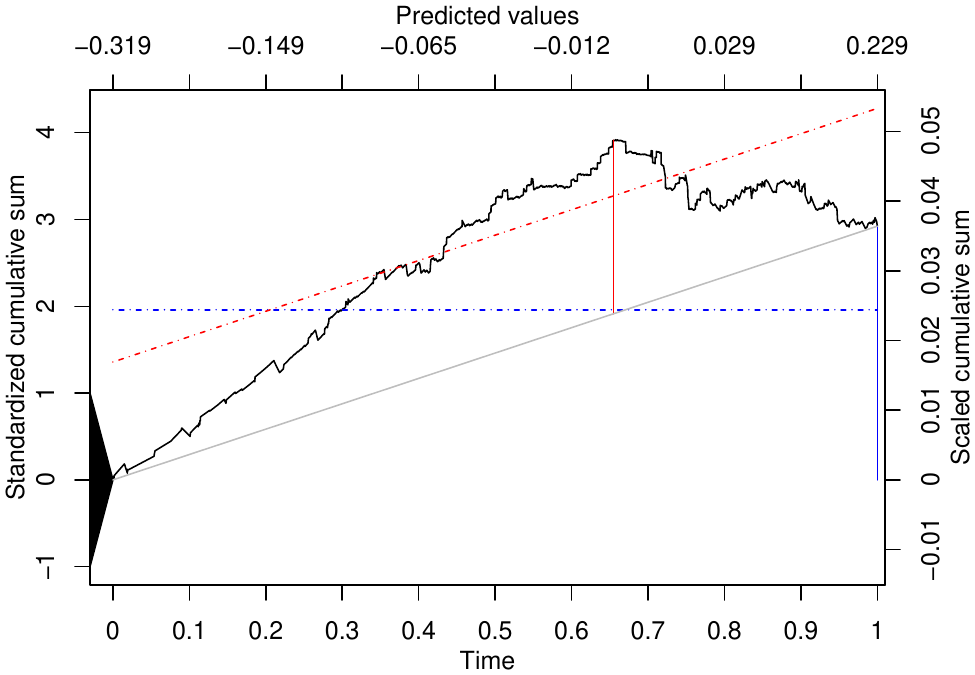}}

\endgroup

\end{document}